\documentclass[manuscript,screen]{acmart}
\usepackage{url}
\usepackage{multirow}
\usepackage{subfigure}
\usepackage{epsfig}
\usepackage{graphicx}
\usepackage{amsmath}
\usepackage{color}
\usepackage{setspace}
\usepackage{hhline}   
\usepackage{algorithm}
\usepackage{algorithmic}
\usepackage{flushend}
\usepackage{graphics}
\usepackage{tikz}
\usepackage{array}% http://ctan.org/pkg/array
\usepackage{balance}
\definecolor{darkgreen}{RGB}{11,140,21}

\begin{document}

\title{Encoder-Decoder Networks for Analyzing 
Thermal and \\ Power Delivery Networks}
\renewcommand{\shorttitle}{Encoder-Decoder Networks for Analyzing Thermal and PDNs}
\author{Vidya A. Chhabria}
\affiliation{%
  \institution{University of Minnesota}
  \country{USA}
}

\author{Vipul Ahuja}
\affiliation{%
  \institution{Qualcomm Technologies Inc.}
  \country{India}
}

\author{Ashwath Prabhu}
\affiliation{%
  \institution{Qualcomm Technologies Inc.}
  \country{India}
}

\author{Nikhil Patil}
\affiliation{%
 \institution{Qualcomm Technologies Inc.}
 \country{India}
 }

\author{Palkesh Jain}
\affiliation{%
  \institution{Qualcomm Technologies Inc.}
  \country{India}
}

\author{Sachin~S.~Sapatnekar}
\affiliation{%
  \institution{University of Minnesota}
  \country{USA}
}

\renewcommand{\shortauthors}{V. A. Chhabria, et. al.}

%\author{Vidya A. Chhabria$^1$, Vipul Ahuja$^2$, Ashwath Prabhu$^2$,  Nikhil Patil$^2$, Palkesh~Jain$^2$, and Sachin~S.~Sapatnekar$^1$ \\ $^1$University of Minnesota, USA;  $^2$Qualcomm Technologies Inc., India}

%\renewcommand{\shortauthors}

\begin{abstract}  % DONE
Power delivery network (PDN) analysis and thermal analysis are computationally expensive tasks that are essential for successful IC design. Algorithmically, both these analyses have similar computational structure and complexity as they involve the solution to a partial differential equation of the same form. 
This paper converts these analyses into image-to-image and sequence-to-sequence translation tasks, which allows leveraging a class of machine learning models with an encoder-decoder-based generative (EDGe) architecture to address the time-intensive nature of these tasks. For PDN analysis, we propose two networks: (i)~IREDGe: a full-chip static and dynamic IR drop predictor and (ii)~EMEDGe: electromigration (EM) hotspot classifier based on input power, power grid distribution, and power pad distribution patterns. For thermal analysis, we propose ThermEDGe, a full-chip static and dynamic temperature estimator based on input power distribution patterns for thermal analysis. These networks are transferable across designs synthesized within the same technology and packing solution. The networks predict on-chip IR drop, EM hotspot locations, and temperature in milliseconds with negligibly small errors against commercial tools requiring several hours.
\end{abstract}

\maketitle

%\bstctlcite{IEEEexample:BSTcontrol}
%\input{sec/outline.tex}

\section{Introduction}
\label{sec:intro}

\subsection{Problem statement}

\noindent
As a consequence of aggressive technology scaling, on-chip power density has been on an increasing trend. This has led to major challenges in designing integrated circuits (ICs) in advanced technology nodes. For successful IC design, it is extremely critical to build tools that can analyze the impact of high power densities. 
Fast and accurate analysis that aids quick design
turnaround is particularly significant for two critical and time-consuming simulations that are performed
several times during the design cycle:

\begin{itemize}
\item {\em Thermal analysis}, which checks the feasibility of a
placement/floorplan solution by computing on-chip temperature distributions to check for temperature hot spots.
\item {\em Power distribution network (PDN) analysis}, which
diagnoses the goodness and reliability of the PDN by checking (i)~whether the voltage (IR) drops from the
power pads to the gates are within specified limits, and (ii)~whether the wire current densities satisfy reliability constraints related to electromigration (EM).
\end{itemize}

However, one of the major challenges with these analyses is the overhead of extremely large runtimes. 
The underlying computational engines that form the crux of both analyses are
similar: both simulate networks of conductances and current/voltage
sources by solving a large system of equations of the form 
$G {\bf V} = {\bf J}$~\cite{Zhan08,Zhong05} 
with millions to billions of
variables.  In modern
industry designs, a single full-chip temperature or IR drop simulation, can take several
hours.
Accelerating these analyses opens the door to optimizations that
iteratively invoke these engines under the hood. 

Prior acceleration methods such as~\cite{multigrid} trade off accuracy for speed by increasing the coarseness of element discretization.
However, the advent of machine learning (ML) has presented fast and accurate solutions to these
problems~\cite{zhang18, juan12, tan19, Lin18,incpird, powernet, thermgan, Wen20}, trained on simulation data from systems of millions of nodes. 

For thermal analysis, existing ML-based solutions primarily focus on coarser system-level thermal modeling~\cite{zhang18,juan12, tan20, thermgan}. The work in~\cite{Wen20} predicts temperature at a finer granularity but uses coarse temperature estimates as an input. The methods in~\cite{tan20,thermgan} use post-silicon performance metrics as inputs to predict the full-chip temperature and are not suitable for predicting temperature during the design process.
%\redfn{There is an inconsistency here: for PDN and EM, you talk about how your work is better than prior work. No such statement for thermal. Can you add a short description, or otherwise find some way to make these consistent? \blueHL{VAC: There seems to be a lot of new papers on this over the last year. Some are journal extensions of the papers we cite in the conference paper. But nothing is really qualitatively directly applicable or comparable as we do for PDN and EM. These are always on the system/package level or specific to microprocessors. Do not delete this comment as it is not addressed. }}

For PDN analysis, the works in~\cite{Lin18, incpird} address
incremental analysis and are not intended for full-chip
estimation.  The work in~\cite{MAVIREC_date21} addresses the vectored IR drop problem and uses ML-based IR drop analysis techniques to recommend a small subset of vectors that exercises worst-case scenarios for dynamic IR drop. 
The work in~\cite{powernet} proposes
a convolutional neural network (CNN)-based implementation for full-chip IR drop
prediction, using cell-level power maps as features. However, it assumes
similar resistance from each cell to the power pads, which may not be valid for practical power
grids with irregular grid density.  The ML techniques in~\cite{powernet,Wen20} both divide the chip into regions ({\it
tiles}), and the CNNs operate on each tile and its near
neighbors.  Selecting an appropriate tile and window size is nontrivial -- small windows could violate the principle of
locality~\cite{Chiprout04}, causing inaccuracies, while large windows could
result in large models with significant runtimes for training and inference. 
Several algorithms~\cite{powernet,MAVIREC_date21} create simplifications that do not account for PDN topologies and power bump locations that are vital for accurate IR drop prediction: as shown in Section~\ref{sec:C4bump_PDNdensity}, this can lead to significant inaccuracies.
Our approach bypasses 
window size selection by providing the entire power map
as input, allowing ML to learn the window size for accurate estimation of temperature and IR drop. 
%While the model in~\cite{MAVIREC_date21} also bypasses the window size selection and predicts full-chip IR drop in a single inference on a per-instance basis, it does not account for PDN topologies and power bump locations that are vital for accurate IR drop prediction.

Our work also considers the use of ML for EM hotspot detection in PDNs. There has been little prior work in this area. The work in~\cite{em-gan} uses a GAN to predict transient stress profiles for multisegment interconnect tree topologies by using images of the tree topologies and their current densities as input. While it is not directly comparable with our work, since it uses multiphysics models as against the empirical models in our work, it is instructive to perform a qualitative comparison. In~\cite{em-gan}, the images are of a fixed $256 \mu$m $\times 256\mu$m size and represent only the small region of a larger chip and interconnect structure. This approach has three limitations. First, the model operates on small regions ($256 \mu$m $\times 256\mu$m) of the chip individually, i.e., if an interconnect belongs to multiple regions, the models predicts stress locally in the region unaware of the rest of the interconnect topology in the neighboring regions. The predicted stress profiles across regions may not maintain continuity across region boundaries when the parts of the interconnect in different regions are stitched back together. Second, PDNs in advanced technology nodes are dense and require a much smaller region size to represent the PDN topologies within the image. However, smaller regions sizes result in scalability and larger inaccuracy issues during stitching. Third, to predict the stress profile, the work uses current density per PDN segment as an input feature. It is computationally expensive to find the current densities in each segment as it requires a solution to the system of equations $G {\bf V} = {\bf J}$. While an ML approach can be used to solve this system of equations rapidly to obtain voltages, this may still be an a linear time solution as it would require iterating through all nodes to generate current densities on a per-segment basis. In addition, most ML methods that solve these equations for IR drop predict IR drop maps at a pixel level, rather than the voltage at each node that is essential for EM analysis at the PDN segment level. 

Our approach overcomes the above limitations by performing EM hotspot classification using a single inference on the entire chip, instead of smaller regions. It encapsulates the estimation of current densities on a per-segment basis into a one-time training step and uses the on-chip power, PDN density distributions, power pad locations, and temperature as inputs to directly predict EM-prone segments during inference.

\subsection{Overview of our approach}

\noindent
In this work, we translate the static analysis problems of thermal, IR drop, and EM hotspot classification into image-to-image translation tasks, and the corresponding dynamic analysis problems into sequence-to-sequence translation tasks. We employ fully convolutional (FC) EDGe
networks for rapid and accurate thermal and PDN analysis for ML image and sequence-based translation tasks. FC EDGe networks
have proven to be very successful with image-related problems with 2D
spatially distributed data~\cite{fcn, unet, mao16, segnet} when compared to
other networks that operate without spatial correlation
awareness.  For transient thermal analysis where the time constant of changes is large, we use long-short-term-memory (LSTM) based networks that maintain a memory of analyses at previous time steps. For transient IR drop, we employ a U-Net based model to capture high-frequency waveforms. Based on these concepts, we propose three novel ML-based analyzers: 

\noindent
{\em ThermEDGe}: \textit{Full-chip} static and transient thermal analysis
    \begin{itemize}
        \item Inputs: Power distributions
        \item Outputs: Temperature distributions
    \end{itemize}
{\em IREDGe}: \textit{Full-chip} static and transient IR drop analysis
    \begin{itemize}
        \item Inputs: Power distributions, PDN density map, and power bump patterns
        \item Outputs: IR drop distributions
    \end{itemize}
{\em EMEDGe}: EM hotspot classification
    \begin{itemize}
        \item Inputs: Power distributions, PDN density map, power bump patterns, and temperature distributions
        \item Outputs: Per PDN layer EM-prone PDN segments
    \end{itemize}

\noindent
The predicted output temperature and IR drop contours are at the accuracy level of a million-node ground truth simulation. The fast inference times of these methods enable {\bf full-chip} thermal, IR drop analysis, and EM hotspot classification in milliseconds, as opposed to several hours using commercial tools. We obtain average errors of 0.6\%, 0.008\%, and 8.5\% (false positives) for ThermEDGe, IREDGe, and EMEDGe, respectively, over a range of testcases.

\begin{figure}[tb]
\centering
\includegraphics[width=0.75\textwidth]{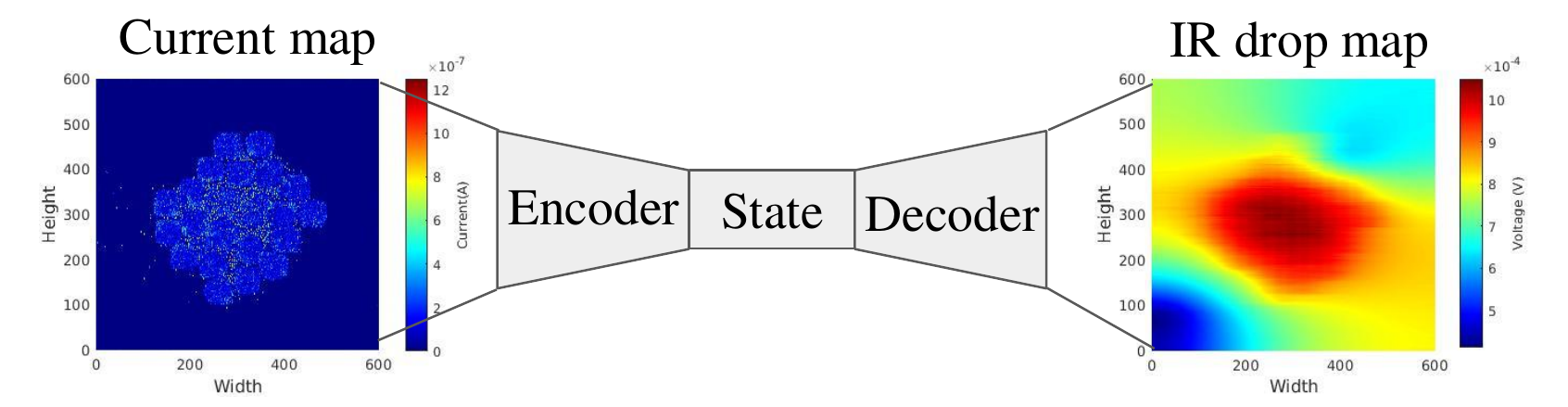}
\caption{Image-to-image translation using EDGe network.}
\label{fig:enc-dec-arch}
\vspace{-1.0em}
\end{figure}

Fig.~\ref{fig:enc-dec-arch} shows a general structure of an EDGe
network. It consists of two parts: (i) the encoder/downsampling path, which
captures global features of the 2-D distributions of power dissipation, and
produces a low-dimensional state space and (ii) the decoder/upsampling path,
which transforms the state space into the required detailed outputs
(temperature or IR drop contours). 
The EDGe network is well-suited for PDN/thermal
analyses because:

\begin{enumerate}
\item[(a)] The convolutional nature of the encoder 
captures the dependence of both problems on the {\it spatial distributions of power}.
Unlike CNNs, EDGe networks contain a decoder that acts as
a generator to convert the extracted power and PDN density features into
accurate high-dimensional full-chip temperature and IR drop contours.
\item[(b)] The trained EDGe network model for static analysis is
{\it design-independent}: it only stores the weights of the convolutional kernel.
Moreover, the same filter can be applied to {\it any} chip of any size for a given technology and packaging solution. The selection of the
network topology (convolution filter size, number of convolution layers) is
related to the expected sizes of the hotspots rather than the size of the chip:
these sizes are generally similar for a given application domain, technology,
and packaging choice.
\item[(c)] Unlike prior methods~\cite{powernet,em-gan} that operate tile-by-tile, where finding the right tile and window size for accurate analysis
is challenging, {\it the choice of window size is treated as an ML
hyperparameter tuning problem} to decide the amount of spatial input information.
\end{enumerate}

This work represents the development and maturation of a previous conference publication~\cite{chhabria21} that addressed the static and transient thermal analysis problems and static IR drop analysis problem using encoder decoder based generative (EDGe) networks.  We extend this to demonstrate the use of EDGe networks for transient IR drop analysis and the EM hotspot classification. 
The organization of the paper is as follows: Section~\ref{sec:prob-def} discusses each analysis in a traditional sense and formulates each analysis into an ML-solvable problem. Section~\ref{sec:edge} justifies the use of EDGe network architecture for these analyses. Section~\ref{sec:train} describes golden data generation and the supervised ML model training details. Section~\ref{sec:results} evaluates EDGe networks in terms of speed and accuracy compared to their golden solvers and prior art counterparts.  

\section{Adapting thermal and PDN analysis to ML}
\label{sec:prob-def}
\noindent
%This paper focuses on addressing three problems by developing three novel tools: (i) ThermEDGE for static and transient %thermal analysis, (ii) IREDGe for static and transient IR drop analysis, and (iii) EMEDGe for EM hotspot classification. 
%While thermal analysis and PDN (EM and IR) analysis are targeted at solving two different problems in chip design, 
The core underlying computations required for PDN analysis (for both IR drop and EM) and thermal analysis are fundamentally similar, solving partial differential equations (PDE) of similar form. 
This section highlights  the conversion of these conventional VLSI-world analyses into ML-world tasks. 

\subsection{Static and transient thermal analysis}
\label{sec:background-therm}

\noindent
The on-chip temperature distribution is governed by the parabolic PDE:
 \begin{equation}
 \label{eq:heat}
         k_t \nabla ^2 T+ g({\bf r}, t) = \rho c_p\frac{\partial T( {\bf r},t)}{\partial t} 
    \end{equation}
    
\noindent
Here, {$\bf r$} is the spatial coordinate of the point at which temperature is being analyzed, $t$ is time (in seconds), $g$ is the power density per unit volume (in W$/$m$^3$), $c_p$ is the heat capacity of the chip material (in J$/$kg K), and $\rho$ is the density of the chip material (in kg$/$m$^3$). Therefore, finding an on-chip temperature profile involves solving $T(r,t)$ given a power density distribution $g(r,t)$ of the chip.

Traditional techniques for solving~\eqref{eq:heat} in either steady-state (i.e., with $\partial T( {\bf r},t)/\partial t = 0$) or transient state are based on the finite difference method (FDM) or finite element method (FEM), which discretize the differential operator or the temperature field across space and time. In steady-state, 
the solution to the above PDE amounts to solving a system of linear equations of the form $G {\bf T} = {\bf P}$~\cite{Zhan08,Zhong05} where $G$ is a $N \times N$ conductance matrix representing connected conductances on the grid, ${\bf T}$ is a $N \times 1$ vector of unknown temperatures, and ${\bf P}$ is a vector of the generated power density values for each element. While $G$ is sparse, in industry-sized designs $N$ is in the order of tens of millions. As a result, on-chip temperature evaluation is computationally expensive even for the static case.

\begin{figure}[tb]
\centering
\includegraphics[width=9.5cm]{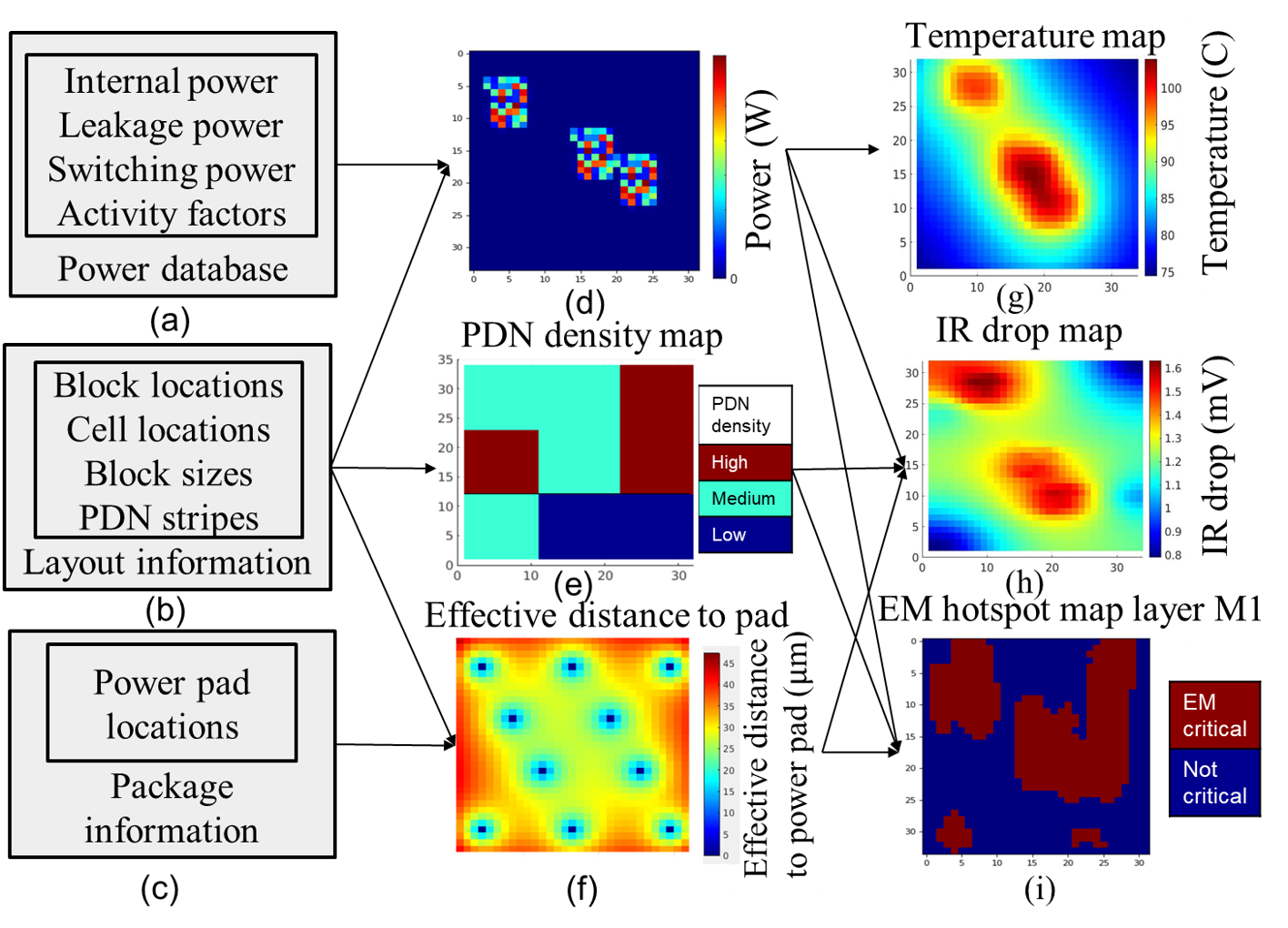}
\caption{Data representation: Mapping PDN and thermal analysis problems into ML-based image-to-image translations tasks. }
\label{fig:data-rep}
\vspace{-2.0em}
\end{figure}

In this work, we translate the above problem of finding $T(r, t)$ for an input $g(r, t)$ given a fixed package ($k_t$, $\rho$, $c_p$) into an ML-solvable problem by representing on-chip power density and temperature distribution as images. The input to~\eqref{eq:heat}, $g(r, t)$, is represented as a power map/image for the static problem and as a sequence of power maps/video frames for the dynamic problem. The maps are created by using per-instance power values from a  power analysis tool and instance locations from the layout database as outlined in Fig.~\ref{fig:data-rep}(a), (b), and (d). Each pixel in a map represents the attributes of a region on the chip. In the power map, each pixel is the sum of power values of each instance in the region and in the temperature map, each pixel is the maximum temperature of all the FDM nodes in that region. We represent the required output, $T(r, t)$ as an image (for static analysis) as shown in Fig.~\ref{fig:data-rep}(g) or a video (for dynamic analysis). The problem can thus be converted into an image-to-image translation task for static analysis and a sequence-to-sequence translation task for dynamic analysis for a fixed $k_t$, $c_p$, and $\rho$.  

ThermEDGe leverages ML models such as U-Nets~\cite{unet} and LSTMs~\cite{lstm} with an encoder-decoder-like architecture for the image-to-image and sequence-to-sequence translation tasks. The computationally expensive FDM analysis is encapsulated into a one-time cost of training ThermEDGe. For a trained network, a rapid inference predicts temperature values of the chip for any given input power map. 
Details of the models and justification for the selection of these architectures are detailed in Section~\ref{sec:edge}.

\subsection{Static and transient IR drop analysis}
\label{sec:background-ir}
\noindent
The on-chip PDN is responsible for transmitting voltages and currents to each cell in the design. However, due to the parasitics in the PDN, a voltage drop is induced between the power pads and the cells in the design. Large voltage drops in the PDN can hurt chip performance and, in the worst case, its functionality. Consequently, it is essential to introduce checks that verify that worst-case IR drop values in the PDN are within specified limits. 
Simulating the PDN to calculate the IR drop in a PDN requires solving a differential equation obtained through modified nodal analysis:
\begin{equation}
 \label{eq:ir}
    G{\bf V}(t) + C\frac{\partial {\bf V}(t)}{\partial t} = {\bf J}(t) 
\end{equation}
where $G$ is the conduction matrix, $C$ is the diagonal capacitance matrix, and ${\bf V}(t)$ and ${\bf J}(t)$ are vectors of voltages and currents at specific instances of time. In steady-state, this reduces to $G{\bf V}~=~{\bf J}$, similar to its thermal counterpart. 

In PDN analysis, on-chip power grid topology and package bump assignments impact IR drop. Therefore,
in addition to representing ${\bf J}(t)$ as a 2D image or sequence of images as in the thermal analysis tasks, both static and transient IR analysis have two other inputs that must be represented as images:

\noindent
(i)~\underline{\it A PDN density map}: This feature is generated by extracting the average PDN pitch in each region of the chip. For example, when used in conjunction with the PDN styles in \cite{jsingh,OpeNPDN}, where the chip uses regionwise uniform PDNs, the average PDN density in each region, across all metal layers, is provided as an input (Fig.~\ref{fig:data-rep}(e)).

\noindent
(ii)~\underline{\it An effective distance-to-power-pad}: We compute the effective distance of each instance, $d_e$, to $N_p$ power pads as 
$$d_e^{-1} = d_1^{-1} + d_2^{-1} + ...+d_{N_p}^{-1}$$
where $d_i$ is the distance of the $i^{th}$ power pad from the instance. Intuitively, the effective distance metric and the PDN density map together reflect the equivalent resistance of all paths between the instance and the pads, where the effective distance-to-power-pad serves as a proxy for resistance. Fig.~\ref{fig:data-rep}(f) shows a typical ``checkerboard'' power pad layout for flip-chip packages~\cite{checkerboard1, checkerboard2}. 

Therefore, IREDGe takes three inputs represented as images: (a)~${\bf J}(t)$ as power maps, (b)~power pad locations that provide an effective-distance-to-power-pad map, and (c)~PDN topologies as PDN density maps. A single input power map is used to predict a static IR drop across the chip (Fig.~\ref{fig:data-rep}(h)), while a sequence of power maps in a short simulation time is used to predict the worst-case dynamic IR drop of that time period~\cite{MAVIREC_date21, powernet}. The IR drop map is an image representation of the IR drop across the chip where each pixel represents the worst IR drop over all PDN nodes on the lowermost layer in that region, which is connected to the switching devices.
%\footnote{With flip-chip packages the voltage sources are connected through C4 bumps at the topmost PDN metal layer which makes the lowermost PDN metal layer connecting to the current sources (the cells on the chip) the layer with the worst IR drop.} 
In this way, we have mapped the static IR drop problem into an image-to-image translation task, and the dynamic IR drop problem into a sequence-to-sequence translation task. 

\begin{figure}[tb]
\centering
\includegraphics[width=0.6\linewidth]{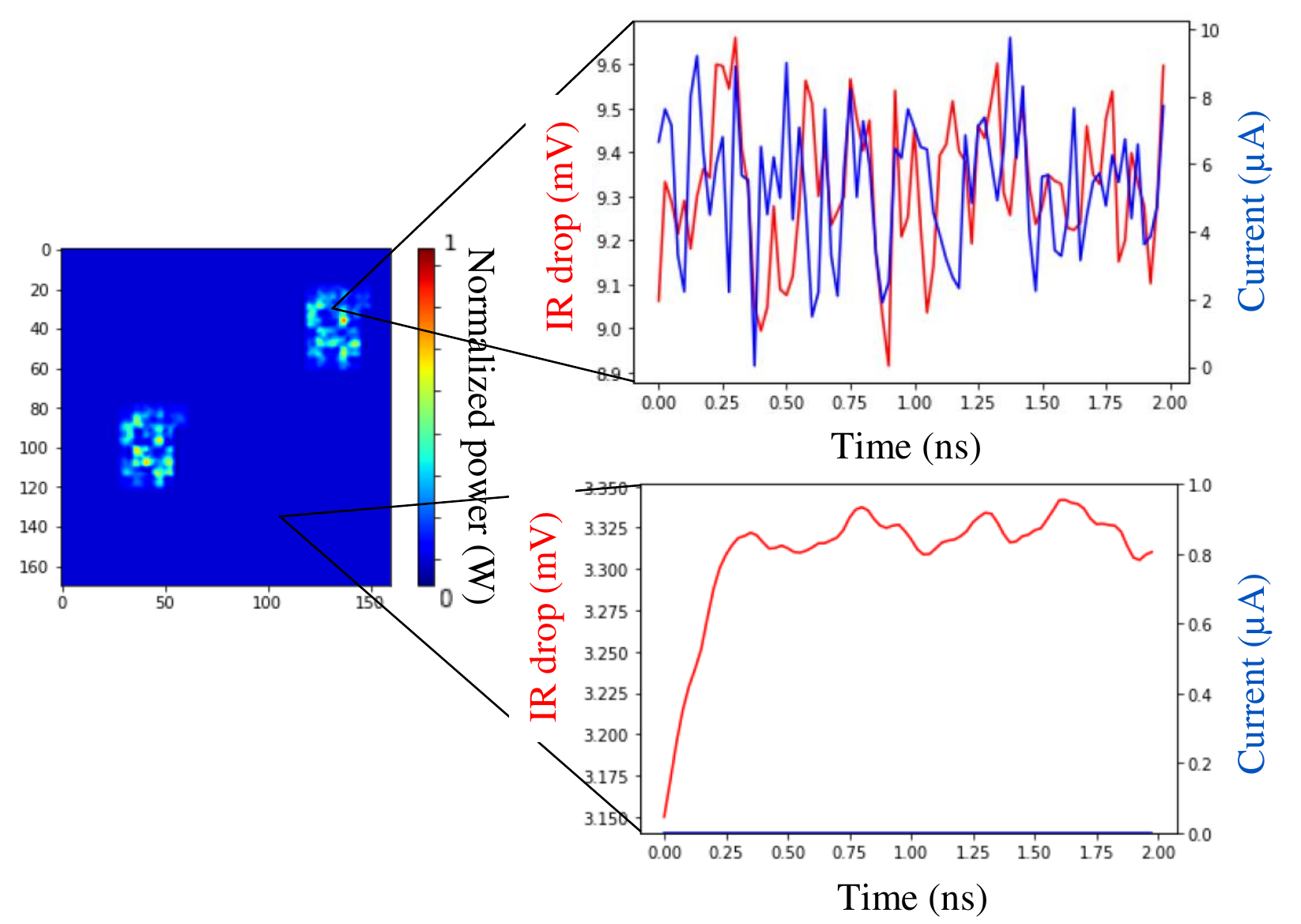}
\caption{High-frequency components of IR drop in regions with high switching activity and low-frequency components of IR drop in regions without activity.}
\label{fig:ir-drop-waveform}
\vspace{-2.0em}
\end{figure}

We use U-Nets for the image-to-image translation task in static IREDGe and a modification of the U-Net that uses 3D convolutional layers for the sequence-to-sequence translation task in transient IREDGe.  Unlike thermal analysis where time constants are large,  IR drop consists of both high-frequency and low-frequency components. The on-chip switching activity contributes the high-frequency components while the package parasitics contribute to the low-frequency components. In our testcases, we observe that regions  with high switching activity are where high-frequency components dominate and regions with low switching activity are where the low-frequency components dominate. This is highlighted in Fig.~\ref{fig:ir-drop-waveform} for a PDN testcase in a 12nm FinFET technology. The high-frequency components contribute to the larger IR drop due to the high switching activity making it essential to capture this accurately. For sequence-to-sequence translation tasks, it has been found in ML literature that 3D convolutional layers better capture local temporal information (high frequency components) than LSTMs (which better capture global temporal information)~\cite{lstmvs3dcnn1, lstmvs3dcnn2} . 
Therefore, 
we use a U-Net model with 3D convolutions as in~\cite{MAVIREC_date21} to predict transient IR drop. Static IR drop is predicted using a U-Net based model as in static ThermEDGe.

% \blue{I still need to justify using sequence-to-image translation task for transient IR and not LSTM networks as in transient thermal analysis. This is probably because of the difference in thermal and IR time constants and the power domain like data we have for thermal analysis. The argument could be made on the basis of the ratio of time constants to switching intervals - higher for thermal than IR - due to which history is less relevant for IR. So an LSTM is not needed if this is true. The catch with this is that transients due to packaging parasitics are typically slower and have higher time constants. See various papers on first droop, for example.  If you have trouble I can dig them up. You may find pointers in, for example, the refs of my ASYNC 2016 paper with Jordi Cortadella. Another thing: in transient analysis, what is your pad model? If you're assuming perfect DC Vdd at the pads, that's not quite realistic, and you won't see first/second/third droop (low-frequency effects). There should be an RLC model for the pads. Doing this will also give you some low-frequency components on top of which the high-frequency components will ride. You could also ping Palkesh to check for realistic modeling. He deals with this on a daily basis.
% }

\subsection{EM hotspot classification}
\label{sec:background-em}
\noindent
EM, a phenomenon caused due to the transport of metal atoms due to the momentum of current-carrying electrons, is a major reliability concern in PDNs.  Empirical EM constraints provided with commercial PDKs dictate that the current density $j$ in each metal segment of the PDN be less that specified limits, which are governed by the Blech filter~\cite{blech:76} and Black's~\cite{black:69} equation.  While there has been recent interest in using accurate multiphysics model-based solutions of Korhonen's equation~\cite{kor:93,TanTahoori19,Shohel21a,Shohel21b}, EM rules in today's  mainstream design technologies do not provide support for these models, and therefore we use empirical models in this work.
%Recent methods that provide fast solutions to Korohonen's equation~\cite{Shohel21a,Shohel21b} could potentially be incorporated into this framework, but are not considered here as they are also not supported by PDKs for contemporary commercial technology nodes.

%The Blech filter eliminates non-critical wires using the product of $j$ and the wire length, and Black's equation determines the mean time to failure of a wire, given by:\redfn{Do we really need this equation or the value of $E_a$? Seems like a diversion. Why not just replace this, someplace in the next paragraph (e.g., after sentence 1) by an equation of the type $I \leq I_{max}$? PDKs don't specifically reference Black's equation or provide $E_a$, afaik.}
%\begin{equation}
%    t_{50} = Aj^{-n}e^{E_a/kT}
%\end{equation}
%Here, $E_a$ = 0.45eV is the activation energy for EM in Cu, $n$
%is the current exponent, and $A$ is a fitting parameter. 
%Therefore, EM analysis requires estimating current density in each segment which amounts to solving~\eqref{eq:ir} for the current in each segment, and involves the same challenges as IR drop analysis in terms of computational complexity. 

In a typical cutting-edge commercial PDK, EM hotspot classification requires comparing the current density in each segment against specified EM limits, i.e., $j \leq j_{max}$, where this limit is modified for shorter wires in the spirit of the Blech criterion. The current density through each segment is obtained by solving~\eqref{eq:ir}. Therefore, estimating the current densities through every segment is also computationally very expensive and runtimes depend on the number of PDN nodes. 

Fig.~\ref{fig:data-rep}(i) shows the EM hotspot map for M1 of the PDN where the red regions indicate EM-critical regions, i.e., at least one PDN segment in that region is EM-critical and the blue segments are non-EM-critical regions. In contemporary technologies, even though the upper metal layers of the PDN carry large unidirectional currents, it is the lower metal layers that are more susceptible to EM failures than upper metal layers due to their small cross-sectional areas.  Therefore, the sizes and number of hotspots decrease as we move to the upper metal layer of the PDN stack.

 Identifying a minority class with a several class imbalance is generally a difficult problem in ML~\cite{class-imbalance-1}. We overcome this using a variety of approaches that are used in the conventional ML literature~\cite{smote, loss-imbalance}. In addition, we use an aggressive EM threshold to successfully train EMEDGe to identify {\it EM-prone} regions and avoid misidentifying an EM-critical segment as EM-safe. The aggressive thresholds also help overcome the class-imbalance (less than 7\% of PDN segments are EM-critical in M5) issue in higher metal layers that are less EM-critical, allowing for successful classification. Therefore, EMEDGe, an ML-assisted EM-prone hotspot identifier, is consistent with the industry preference where a fast ML inference can quickly identify EM-prone PDN segments, and an accurate yet fast small-scale simulation can evaluate the predicted hotspots during chip sign-off.

In this work, we propose using ML to overcome the computationally expensive nature of estimating the current density in each segment by using EMEDGe to predict EM-prone regions directly through a fast inference. We set up EMEDGe as a hotspot classifier that recommends EM-prone regions instead of identifying truly EM-critical segments directly. Since EM requires local change of the power grid density, it is helpful to flag the larger region for PDN resynthesis. 
 In our framework, EMEDGe flags EM-prone regions that can later be analyzed for EM-criticality using a fine-grained accurate simulation of the PDN at a much smaller scale (fewer PDN nodes) than the original problem using hierarchical PDN analysis~\cite{zhao02}. We use an aggressive EM threshold to successfully train EMEDGe to identify EM-prone regions and avoid misidentifying an EM-critical segment as EM-safe. The aggressive thresholds also help overcome the class-imbalance (less than 7\% of PDN segments are EM-critical in M5) issue in higher metal layers that are less EM-critical, allowing for successful classification. Therefore, EMEDGe, an ML-assisted EM-prone hotspot identifier, is consistent with the industry preference where a fast ML inference can quickly identify EM-prone PDN segments, and an accurate yet fast small scale simulation can evaluate the predicted hotspots during chip sign-off.

It is important to note that the voltage predicted by IREDGe cannot be directly used for estimating the current density in each segment for EM checks for two reasons: (i) IREDGe predicts the voltage drop in the lowermost layer of the PDN (M1) and is unaware of the PDN node location and segment level information in other layers. Therefore, we cannot estimate the current density through other layers in the PDN which can potentially be EM-critical, and (ii) even if we modify IREDGe to generate the voltage at each PDN node, using this method would still be slower compared to the direct use of a single ML inference as it would still require one iteration through all PDN nodes.

% In our testcases, the metal layers M1 and M2 have 30-50\% of their PDN segments as EM-critical, M5 has less than 7\% of EM-critical segments, and M8 and M9 have zero EM-critical segments in the dataset. The small number of EM-critical segments in M5 makes it extremely challenging to capture hotspots in this layer due to a severe imbalance in data. To address this, we employ equally aggressive EM thresholds across all layers such that the fraction of the number of EM hotspots in M5 increases from 7 to 15\%. While this threshold scaling does increase the number of false positives on the original threshold, it ensures EMEDGe does not miss reporting any EM-prone regions. The flagged regions can be checked with a more accurate detailed analysis on a much smaller scale (few thousand nodes as compared to millions) than the original problem itself. The application of EMEDGe is in detecting regions with EM-prone segments that can later be analyzed in detail using fast solvers. Therefore, EMEDGe which serves as an ML-assisted EM-hotspot identifier and is consistent with the industry requirements where a fast ML inference can quickly identify EM-prone PDN segments and an accurate yet fast small scale simulation can evaluate the predicted hotspots.

In standard flows, the current densities in each PDN segment are compared against EM limits to determine if a segment is EM-critical. The EM limits are specific to a technology, the BEOL stack, and are a function of the width of the PDN stripe and temperature. EMEDGe operates on a layer basis requiring one model per metal layer used in PDN.
The four spatial inputs, represented as images, to EM hotspot classification are: (a) ${\bf J}$ as a power map, (b) PDN topology as a PDN density map, (c) power pad locations as an effective-distance-to-power-pad map, and (d) and an on-chip temperature map obtained from either the output of ThermEDGe or solving~\eqref{eq:heat}\footnote{We use on-chip temperature as a feature for EM hotspot classification to capture (i) temperature-dependent EM limits and (ii) the exponential dependence of current densities on temperature.}. Unlike static IREDGe, which predicts worst on-chip IR drop on the {\em lowermost layer}, EMEDGe must predict EM hotspots which highlight EM-prone PDN segments for {\em each layer} in the PDN. Therefore, EMEDGe consists of multiple trained models for each layer in the PDN to predict EM hotspots separately. 

EMEDGe converts the EM hotspot classification problem into an image segmentation (image-to-image translation) problem where the goal is to classify each region as either EM-prone or EM-safe. We use a U-Net model to predict the EM hotspot locations. EMEDGe is metal layer-specific, i.e., for a PDN with five metal layers, we use five U-net based models to predict the EM hotspots for that specific layer. The inputs for the five models are identical but the labels are layer-specific.  

\noindent
\underline{Summary}: We summarize the different PDN and thermal analysis problems that we address in this work in Table~\ref{tbl:summary}. The table highlights the differences in inputs, outputs, and the EDGe network used for each analysis. The static analyses are treated as image-to-image translation tasks and use U-Nets the transient analyses are treated as sequence-to-sequence translation tasks and use 3D U-Nets in the case of IR drop analysis and LSTMs for thermal analysis.

\begin{table}
\centering
\caption{Summary of problems addressed with the corresponding EDGe network, their input features, and output labels.}
\label{tbl:summary}
\resizebox{0.9\linewidth}{!}{%
\begin{tabular}{||c|l||l|l||l|l||} 
\hhline{|t:==:t:==:t:==:t|}
\multicolumn{2}{||c||}{\textbf{Analysis type}} & \multicolumn{1}{c|}{\textbf{ML task}} & \multicolumn{1}{c||}{\textbf{EDGe~}} & \multicolumn{1}{c|}{\textbf{Inputs}} & \multicolumn{1}{c||}{\textbf{Output}} \\ 
\hhline{|:==::==::==:|}
\multirow{2}{*}{\begin{tabular}[c]{@{}c@{}}\textbf{Thermal }\\\textbf{analysis}\end{tabular}} & \begin{tabular}[c]{@{}l@{}}\textbf{Static }\\\textbf{thermal}\end{tabular} & \begin{tabular}[c]{@{}l@{}}Image-to-image \\translation\end{tabular} & U-Net~ & Power map & Temperature map \\ 
\cline{2-6}
 & \begin{tabular}[c]{@{}l@{}}\textbf{Dynamic }\\\textbf{thermal}\end{tabular} & \begin{tabular}[c]{@{}l@{}}Sequence-to-sequence \\translation\end{tabular} & LSTM~ & \begin{tabular}[c]{@{}l@{}}Time varying \\power maps\end{tabular} & \begin{tabular}[c]{@{}l@{}}Time varying \\temperature maps\end{tabular} \\ 
\hline
\multirow{3}{*}{\begin{tabular}[c]{@{}c@{}}\textbf{PDN }\\\textbf{analysis}\end{tabular}} & \begin{tabular}[c]{@{}l@{}}\textbf{Static}\\\textbf{IR drop}\end{tabular} & \begin{tabular}[c]{@{}l@{}}Image-to-image \\translation\end{tabular} & U-Net & \multirow{3}{*}{\begin{tabular}[c]{@{}l@{}}Power map \\(time varying for dynamic), \\power pads locations, \\ PDN topology,~and \\ temperature (for EM only)\end{tabular}} & Static IR drop map \\ 
\cline{2-4}\cline{6-6}
 & \begin{tabular}[c]{@{}l@{}}\textbf{Dynamic }\\\textbf{IR drop}\end{tabular} & \begin{tabular}[c]{@{}l@{}}Sequence-to-sequence \\translation\end{tabular} & 3D U-Net &  & \begin{tabular}[c]{@{}l@{}}Worst-case dynamic \\IR drop map\end{tabular} \\ 
\cline{2-4}\cline{6-6}
 & \begin{tabular}[c]{@{}l@{}}\textbf{EM}\\\textbf{hotspot}\end{tabular} & \begin{tabular}[c]{@{}l@{}}Image-to-image \\translation\end{tabular} & U-Net &  & \begin{tabular}[c]{@{}l@{}}EM hotspot map\\per-layer\end{tabular} \\
\hhline{|b:==:b:==:b:==:b|}
\end{tabular}
}
\end{table}

\section{EDGe Networks for Thermal and PDN Analysis}
\label{sec:edge}

\subsection{U-Nets for image-to-image translation}
\label{sec:unet}
\noindent
\subsubsection{Overview of U-Nets}
CNNs are successful in extracting 2D spatial information for image
classification and image labeling tasks, which have low-dimensional outputs (class or label). For PDN and thermal analysis tasks, the required outputs
are high-dimensional distributions of IR drop/EM hotspots and temperature
contours, respectively, where the dimensionality corresponds to the number of
pixels of the image and the number of pixels 
is proportional to the size of the chip. This calls for a generator network
that can translate the extracted low-dimensional input features such as (power, PDN, effective distance to pad features, etc.) from a
CNN-like encoder back into high-dimensional data representing the required output. 

\begin{figure}[htb]
\centering
\includegraphics[width=9.5cm]{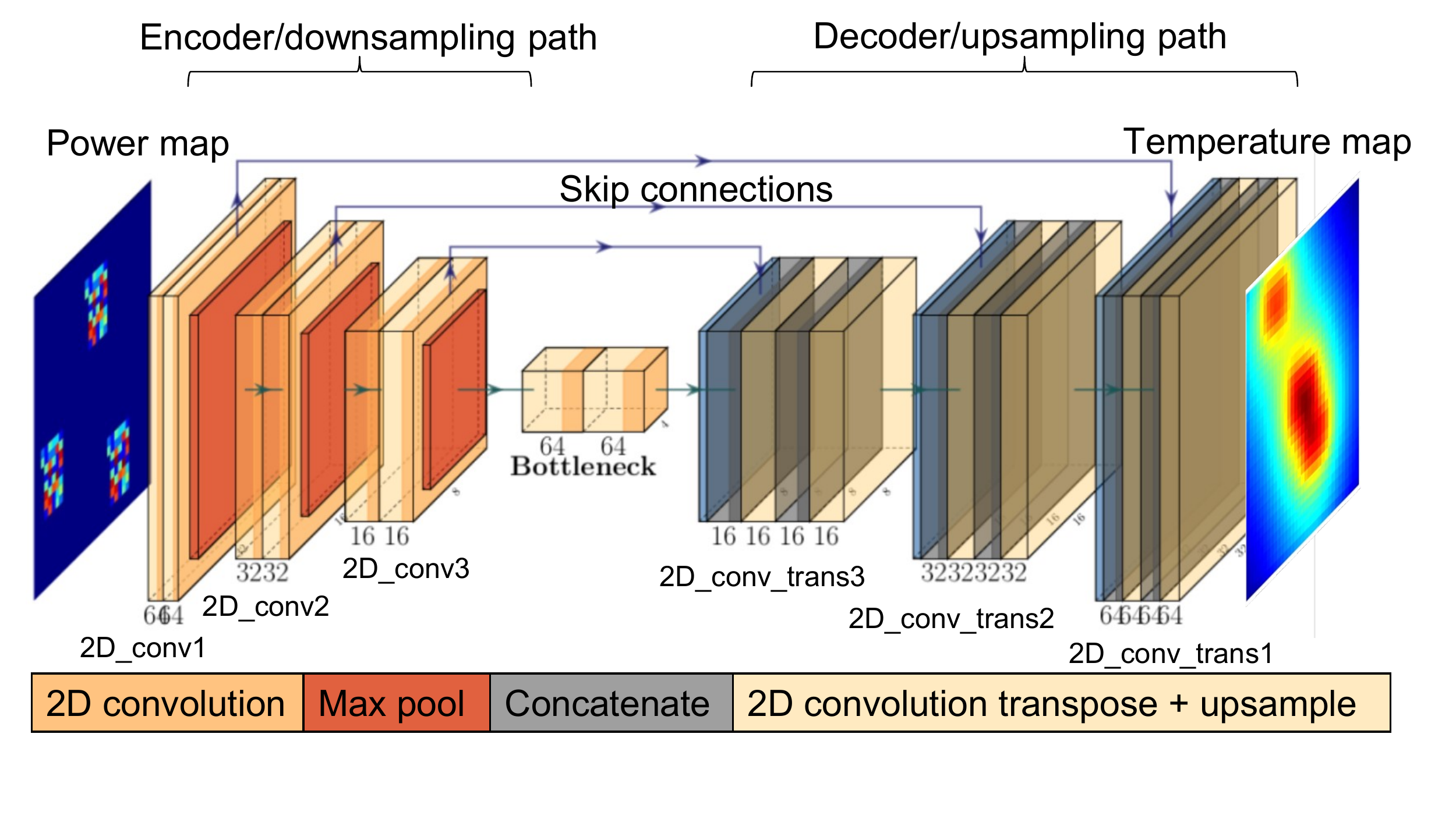}
\vspace{-2.5em}
\caption{U-Net-based static IREDGe, ThermEDGe, and EMEDGe.}
\vspace{-1.0em}
\label{fig:thermedge}
\end{figure}

Fig.~\ref{fig:thermedge} shows the
structure of the EDGe network used for static PDN (IR and EM) and thermal analysis.
It consists of two networks:

\noindent
(a)~\underline{\it Encoder/Downsampling Network} Like a CNN, the network utilizes a sequence of 2D convolution and max pooling layer pairs that extract key features from the
high-dimensional input feature set.  The convolution operation performs a
weighted sum on a sliding window across the image~\cite{conv-deconv}, and the
max pooling layer reduces the dimension of the input data by extracting the
maximum value from a sliding window across the input image. In
Fig.~\ref{fig:thermedge}, the feature dimension is halved at each stage by each
layer pair, and after several such operations, an encoded, low-dimensional,
compressed representation of the input data is obtained. For this reason, the
encoder is also called the downsampling path: intuitively, downsampling
helps understand the {\it ``what"} (e.g., ``Does the image contain power or EM or IR
hotspots?'') in the input image but tends to be imprecise with the {\it ``where"}
information (e.g., the precise locations of the hotspots).  The latter is
recovered by the decoder stages.

\noindent
(b)~\underline{\it Decoder/Upsampling Network} Intuitively, the generative decoder is
responsible for retrieving the {\it ``where"} information that was lost during
downsampling, This distinguishes an EDGe network from its CNN counterpart.  The
decoder is implemented using the transpose convolution~\cite{conv-deconv} and
upsampling layers. Upsampling layers are functionally the opposite of a
pooling layer and increase the dimension of the input data matrix by replicating rows and columns.  

\subsubsection{Use of Skip Connections}

The outputs of the PDN analysis and thermal analysis are  strongly correlated to the input power -- 
a region with high power on the chip could potentially have a IR drop or EM hotpots or temperature hotspots in its vicinity.  U-Nets~\cite{unet} utilize {\it skip} connections between the downsampling and upsampling paths, as shown in
Fig.~\ref{fig:thermedge}.  These connections take information from one layer
and incorporate it using a {\it concatenation} layer at a deeper stage skipping intermediate layers and appends it to the embedding along the z-dimension.  

Skip connections combine the local embeddings  
% power, PDN information, and power pad locations
from the
downsampling path with the global power information from the upsampling path,
allowing the underlying input features to shuttle to the layers closer to the output directly.  This helps recover the fine-grained ({\it ``where"}) details that are lost in the encoder network (as stated before) during upsampling in the decoder for detailed IR drop contours, EM hotspots, and temperature contours.

\subsubsection{Receptive Fields in the Encoder and Decoder Networks}
The characteristic of PDN and thermal analysis problems is that the IR drop, EM current densities, and 
the temperature at each location depends on both the local and global power
information. 
%This is predominant in the thermal analysis where power dissipation in one region of the chip can significantly affect temperature in other parts of the chip due to the large heat diffusion length constants.
The network captures local spatially correlated distributions during convolution by sliding averaging windows of an appropriate size
across the input power image. For capturing the larger global impact of power  on IR drop, EM current densities, and temperature max pooling layers are used after each convolution to
appropriately increase the size of the {\it receptive field} at each stage of
the network. The {\it receptive field} is defined as the region in the input
2D space that affects a particular pixel, and it determines the impact
local, neighboring, and global features have on the analyses.  

In a deep network, the value of each pixel feature is
affected by all of the other pixels in the receptive field at the previous convolution stage, with the largest contributions coming from pixels near the center
of the receptive field.  Thus, each feature not only captures its receptive field in the input image but also gives an exponentially higher weight to the middle of that region~\cite{receptive-field}. This matches
with our applications, where IR, EM, and temperature hotspots maps for a pixel are most
affected by the features in the same pixel, and partially by features in
nearby pixels, with decreasing importance for those that are farther away.
The size of the receptive field at each stage in the network is determined by the filter sizes and the number of the convolutional and max pooling layers. The number of layers and filter sizes is determined based on the magnitude and size of the hotspot that are encountered during design iterations.

%On both the encoder and decoder sides in Fig.~\ref{fig:thermedge}, we use three
%stacked convolution layers, each followed by 2$\times$2 max-pooling to
%extract the features from the power and PDN density images. 

\subsection{3D U-Nets for IR drop sequence-to-sequence translation}
\noindent
U-Nets consist of 2D convolutional layers that perform the convolution operation on the input across {\em all} the input channels of the input features. Due to the averaging nature of the 2D convolution operation, the resulting embedding loses fine-grained detailed information that in each channel. While this averaging effect helps capture spatial variations of power for the static IR drop and static thermal analysis problems, it fails to capture detailed fine-grained information such as the on-chip switching activity that is critical to the transient IR drop problem. Therefore, for the transient IR drop problem, we use a 3D U-Net, inspired by~\cite{MAVIREC_date21}, which uses 3D convolutional layers in the encoder path instead of 2D convolutional layers. The 3D convolutional layer restricts the number of input channels on which the convolutional layer operates to a small local window based on the specified filter size.  In the case of transient analysis IR drop analysis, the input channels represent time-varying power maps where each channel is the power at a specific time-step. Thus, the 3D convolutional layer works on a few time-steps of power instead of all time-steps at once. This prevents the loss of fine-grained switching information due to the averaging effect across all channels. Note that since the decoder path uses 2D convolutional layers, the interface between the 3D embedding in the encoder and 2D embedding is a sum across all channels of the embedding through the skip connections.

\subsection{LSTMs for thermal sequence-to-sequence translation}
\noindent
Long short term memory (LSTM) based EDGe networks are a special kind of recurrent neural
network (RNN) that are known to be capable of learning long term dependencies
in data sequences, i.e., they have a memory component and are capable of learning
from past information in the sequence. 

\begin{figure}[h]
\centering
\includegraphics[width=0.7\textwidth]{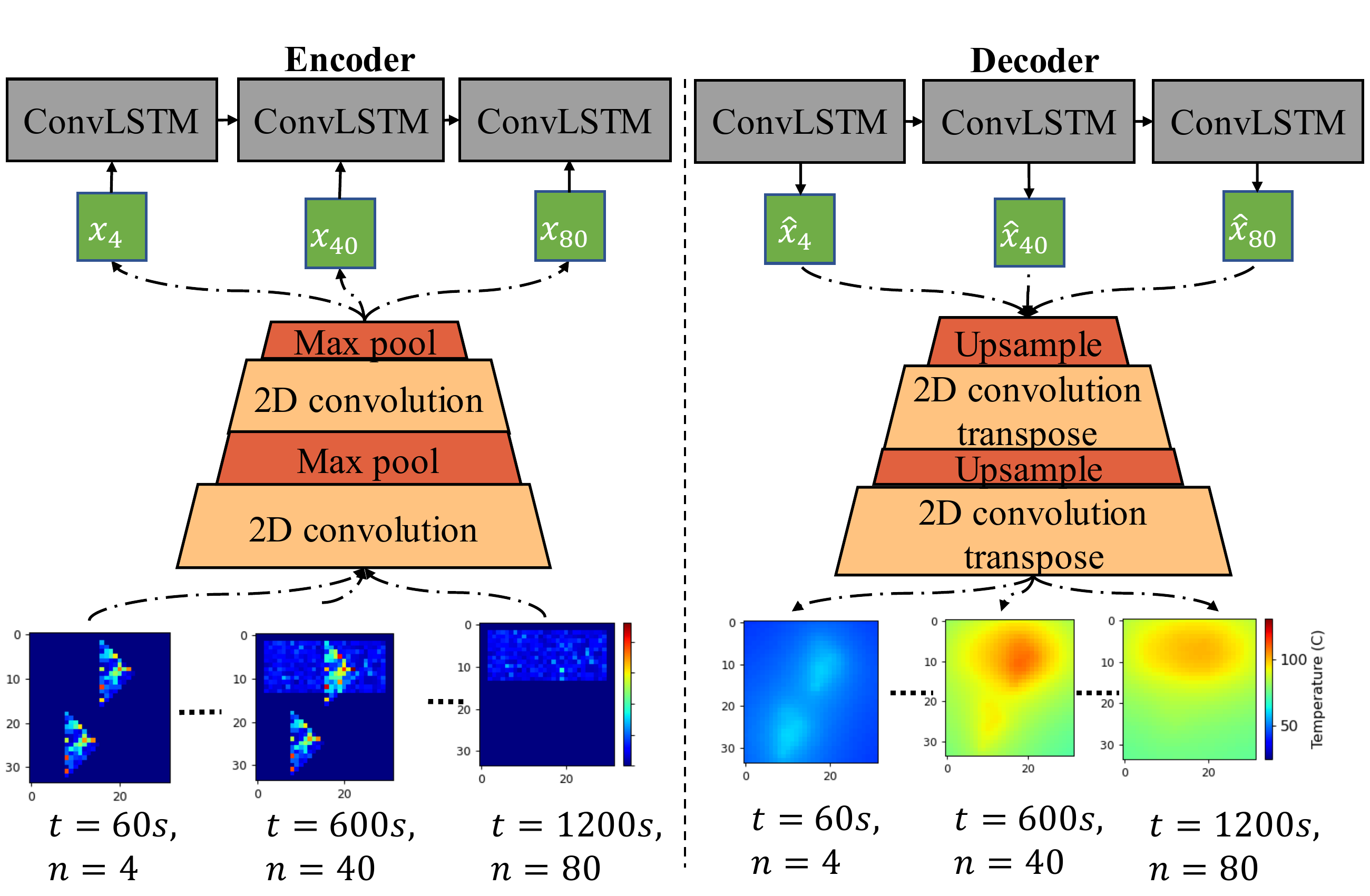}
\caption{LSTM-based network for transient ThermEDGe.}
\vspace{-1.0em}
\label{fig:lstm-edge}
\end{figure}

\begin{figure}[h]
\centering
\includegraphics[width=9cm]{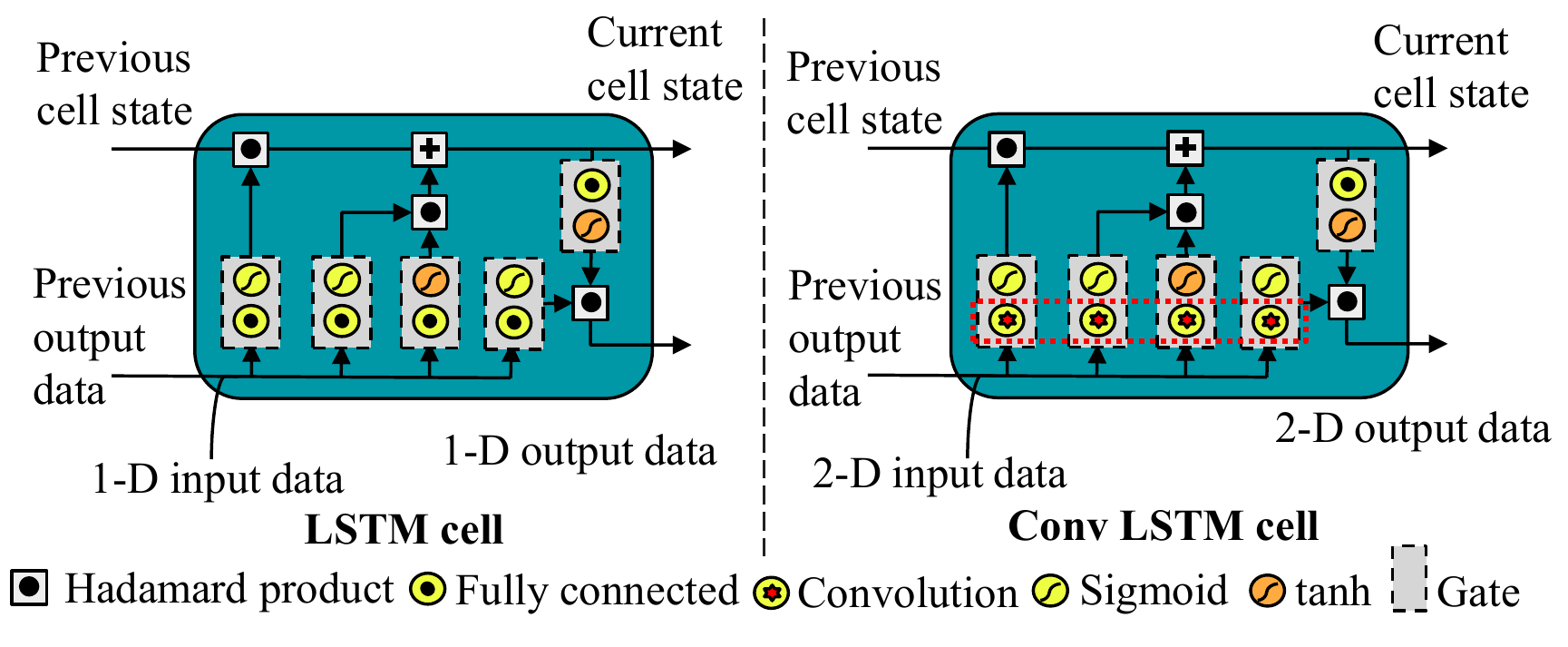}
\caption{Conventional (left) and ConvLSTM (right) cells.}
\label{fig:conv-lstm}
\end{figure}

For transient thermal analysis,  the structure of ThermEDGe is shown in Fig.~\ref{fig:lstm-edge}. The core architecture is an EDGe network, similar to the static analysis problem described in Section~\ref{sec:unet}, except that the 
network uses additional LSTM cells to account for the time-varying component. The figure demonstrates the time-unrolled LSTM where input power
frames are passed to the network one frame at a time. The LSTM cell accounts
for the history of the power maps to generate the output temperature frames.  The network is used for sequence-to-sequence translation in transient thermal analysis, where the input is a set of time-varying power maps and the output is a set of time-varying temperature maps (Section~\ref{sec:prob-def}).

Similar to the static ThermEDGe network (Fig.~\ref{fig:thermedge}), the encoder consists of convolution and max pooling layers to downsample and extract critical local and global spatial information, and the decoder consists of upsampling and transpose convolution layers to upsample the encoded output. 
However, in addition, transient ThermEDGe has LSTM layers in both the encoder and decoder.

A standard LSTM cell is shown in Fig.~\ref{fig:conv-lstm} (left). 
% The key element is the cell state, represented by the
% horizontal line running through the top of the cell.  The state line,
% which depends on information from the previous output data, interacts with
% gates that determine how the current inputs will influence the cell state, and
% preserves a ``memory'' over the sequence.  The sigmoid layer within the gate
% outputs a value between zero and one, determining how much of the input should
% be transmitted through the gate.
While the basic LSTM cell uses fully connected layers within each gate, our application
uses a variation of an LSTM cell called a convolutional LSTM
(ConvLSTM)~\cite{conv-lstm}, shown in Fig.~\ref{fig:conv-lstm} (right).  In
this cell, the fully connected layers in each gate are replaced by convolution
layers that capture spatial information.  Thus, the LSTM-based EDGe network
obtains a spatiotemporal view that enables accurate inference.

\section{EDGe Network Training}
\label{sec:train}
\noindent
We train ThermEDGE, IREDGe, and EMEDGe to learn temperature, IR drop contours, and EM hotspots. The training process consists of ``golden'' data generation and ML model training which are explained in detail in the rest of this section.

% \blueHL{The outline for this section is essentially as follows. Delete once written.
% \begin{enumerate}
%     \item Golden data generation: For each tool answer the following questions:
%         \begin{enumerate}
%             \item Which tool is used for data generation?
%             \item How many data points are generated?
%             \item How long does it take to generate one datapoint?
%             \item What is the training specific to? Size, layer, hotspot size,
%         \end{enumerate}
%     \item Model training
%         \begin{enumerate}
%             \item Model architectures EMEDGe, IREDGe, and ThermEDGe
%             \item Training hyperparameters
%         \end{enumerate}
% \end{enumerate}
% }

\subsection{Generating Training Data}
\label{sec:train-data}
\noindent
A challenge that we faced while evaluating our techniques is the dearth of public domain benchmarks that fit these applications. The IBM benchmarks~\cite{ibm}, are potential candidates for our applications, but they assume constant currents per region and represent an older technology node. Therefore, we generate our 
dataset, which comprises of industry-relevant testcases, where each testcase represents industry-standard workloads for commercial designs implemented in FinFET technology. 

For all three models, ThermEDGe, IREDGe, and EMEDGe, we use images/maps of size 34$\times$32 pixels where each pixel represents a 250$\mu$m$\times$250$\mu$m resolution tile for thermal analysis and 25$\mu$m$\times$25$\mu$m resolution tile for PDN analysis.\footnote{Note that although the temperature, IR drop, and power maps work at this resolution, the ground-truth simulation consists of millions of nodes; using fewer nodes (e.g., one node per pixel) is grossly insufficient for accuracy.}  This difference in resolution is required because thermal hot-spots have a larger spatial spread than IR hot-spots.
%~\redHL{Note that this means the chip sizes are different. For PDN we use chip sizes of 0.68mm2 and for thermal we use chip sizes of 68mm2. This is due to scalability issues in PDNSim. So feel free to change the location of this if you think its not going to please the reviewers very much.}\redfn{I assume this red text is a note to me and not content to be listed in the paper? If so, I'm ok with deleting it all. If you intend to keep it, pl. let me know which parts you feel are important to mention. The bigger problem is that this comes out of the blue: why does thermal and IR require a different spatial resolution? Would be good to explain in some way (e.g., IR decays more rapidly in space?) \blueHL{ Yes. Red text for you. This does not have to do with the spatial decay of IR drop. This is essentially how much of a region size the one pixel represents. So I have testcases from Palkesh et. al. that are power maps with 34x32 pixels. I need to scale this back to a real chip size to synthesize the power grid. So for scalability purposes of PDNSim, I assume the 34x32 power map represents a chip of area 0.68mm2. This chip already has over 3M nodes.  However, in reality, as per the testcase provided it is a 68mm2 chip. For thermal analysis I already have the golden data from Icepak and do not run into scalability issues of PDNSim. This is important since we are working with two different chips essentially for PDN and thermal analysis although the power distribution is similar.}}  
We reiterate that although the training is performed on chips of fixed size, as
we show (Section~\ref{sec:results}), inference can be performed on a chip of any size as long as the resolution (250$\mu$m$\times$250$\mu$m or 25$\mu$m$\times$25$\mu$m) and technology remain the same.

\subsubsection{Thermal Analysis}
For both the static and thermal analysis problems, we obtain our golden data from Ansys-Icepak~\cite{icepak} simulations. Each simulation is expensive in terms of time and memory resources: one simulation of static thermal analysis takes about 40 minutes and one simulation of transient thermal analysis for a 3000s time interval with 45 time-steps takes over 4 hours with $\approx 2$ million nodes. 

We create a 50-datapoint set for the static thermal analysis problem, where a single datapoint consists of a power map and static temperature maps. For the transient thermal analysis case, our training data consists of 150 datapoints with time-varying workloads as features and the
time-varying temperatures as labels.
For each testcase, we generate 45 time-step simulations that range from 0 to
3000s, with irregular time intervals from the thermal simulator. The LSTM-based network is trained using constant time steps of 15s, which enables
easy integration with existing LSTM architectures, which have an implicit
assumption of uniformly distributed time steps, without requiring additional
features to account for the time.

\subsubsection{IR drop analysis} 
We synthesize irregular PDNs of
varying densities for each dataset element using {\it PDN templates}, as
defined by OpeNPDN~\cite{OpeNPDN}. These templates are a set of PDN building
blocks, spanning multiple metal layers in a 12nm commercial FinFET technology, which vary in metal
utilization. For our testcases, we use three templates (high, medium, and low
density) and divide the chip into nine regions. As outlined in Section~\ref{sec:background-ir}, we use a checkerboard pattern of power pads that vary in the bump pitch and offsets across the dataset.
%As a side note, a checkerboard pattern is a common VDD/VSS bump assignment pattern in flip-chip packages with regular rectangular bump arrays~\cite{checkerboard1}.

For static analysis, we create a 5000-datapoint set with a combination of 50 different power distributions, 10 different PDN densities, and 10 different patterns of power pad distributions. Unlike thermal analysis which has only power as the input feature, IR drop analysis as three additional input features (Refer~\ref{tbl:summary}) as degrees of freedom requiring a larger dataset to successfully train the model. For both the static and transient PDNs and power maps, we use an open-source modified nodal analysis-based IR drop solver PDNSim~\cite{pdnsim} to generate ``golden" datapoints\footnote{The open-source version of PDNSim is for static IR drop analysis only. Therefore, we modify the source code to perform dynamic IR drop analysis as in~\eqref{eq:ir}.}. A single static IR drop simulation for a PDN with 2M nodes takes 20 minutes, and a single transient IR drop simulation for a 2ns duration takes $\approx$40 minutes.  The simulation generates the voltage drop at every PDN node. We use the voltages to create an IR drop map represented at the same resolution as the input power map, where each pixel represents the maximum IR drop of all PDN nodes in a 25$\mu$m$\times$25$\mu$m region.

\begin{figure}[htb]
\centering
\includegraphics[width=9.2cm]{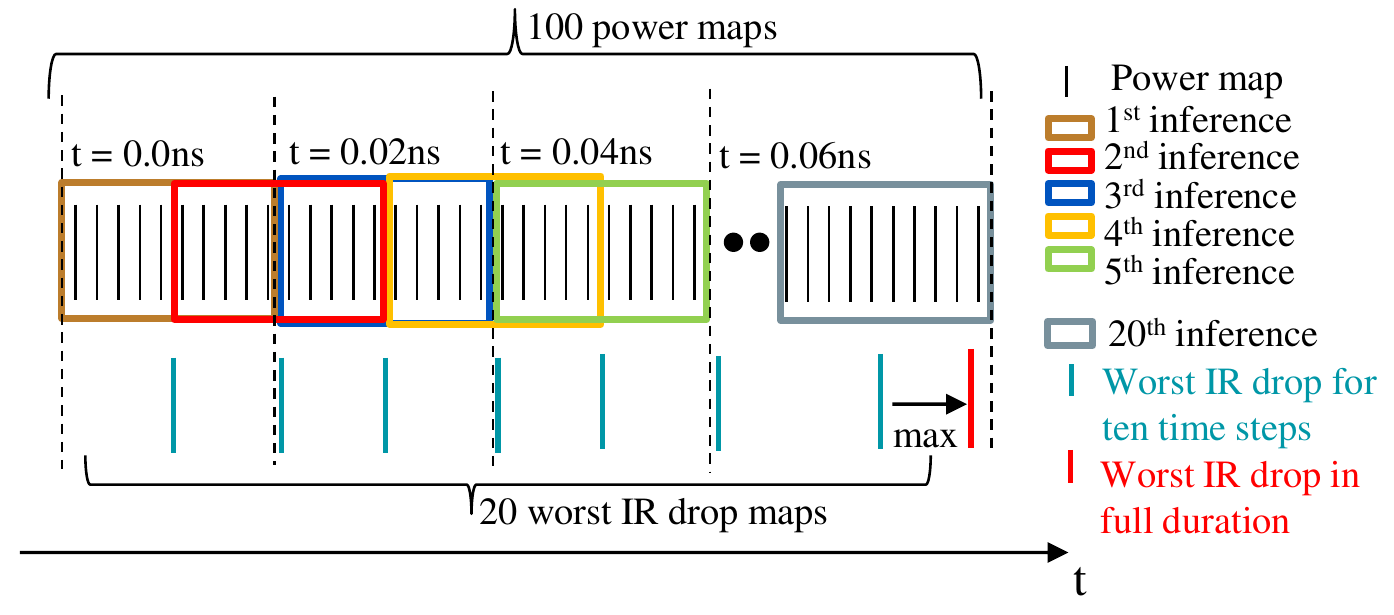}
%\vspace{-2.5em}
\caption{Transient IREDGe power map sequence used for training and inference. The U-Net based model with 3D convolutions predicts IR drop for each inference.}
\label{fig:trans-ir-time-steps}
\end{figure}

Due to the dearth of modern public domain transient PDN analysis benchmarks, we use the benchmarks for static analysis and add randomly generated time-varying components to the power-consuming areas of the chip to generate our transient benchmarks. These components model the switching activity within the current power mode to generate time-varying power distributions. We generate 100 time-step power maps where each time step represents one-hundredth of a 2ns clock period. 

The U-Net is trained to predict the worst IR drop for an overlapping sequence of ten time-steps. Although we use a 3D convolution layers that work on a small local window of neighboring time-steps we still use an overlapping sequence of time-steps for training and inference. The overlapping sequence provides the model neighboring past and future switching activity to accurately predict the worst IR drop for the current sequence. The number of time steps and number of overlapping time steps considered during training are ML hyperparamters that are tuned for best accuracy.
 Fig.~\ref{fig:trans-ir-time-steps} illustrates how transient IREDGe uses the sequence of power maps. The figure shows that we predict the worst IR drop for every ten time-steps in the sequence of power maps but use an overlapping five-time-step sequence of power maps. Therefore, to predict the IR drop for a $t=2$ns duration, we perform 20 inferences.  
We find that working at this finer granularity with multiple inferences, where each inference predicts the worst IR drop for fraction of the clock period, provides better accuracy than using all the 100 time-steps of the input power maps with a single inference.

\subsubsection{EM hotspot classification}
While PDNSim~\cite{pdnsim} generates the voltage at every PDN node, the ``golden" data for training EMEDGe requires estimating the current in each segment and checking it against the PDK-specified EM limits for that segment. We use PDNSim to generate the current densities through each segment given power distribution, power pad bump distribution, and temperature distributions for each testcase and compare against the technology-specific EM limits.  The limits are set based on process design kit (PDK) requirements for a commercial 12nm FinFET technology. The PDK specifies limits as a function of lifetime, temperature, layer, and length for a given width. In our work, we use a 10 year lifetime and derate the limit for each PDN segment based on the input temperature using the limit characterization from the PDK and flag each segment in the PDN as EM-critical or not.  We convert the per-segment information into an EM hotspot map, which has the same resolution as the input images, by annotating a pixel as critical if any PDN segment in the region represented by the pixel has an EM-critical segment. The runtimes for generating a single data point in the training dataset for EMEDGe are identical to those of the static IREDGe simulations.  

Similar to IREDGe, our training datasets for EMEDGe consists of 5000 datapoints for each of the five models (for the five PDN layers) which is generated by a combination of 50 different power distributions, 10 different PDN densities, 10 different patterns of power pad distributions, and different temperature combinations. Unlike IR drop analysis, where we train a single model to predict the on-chip IR drop, for the EM hotspot classification problem, we predict EM hotspots for every layer in the PDN.  For the PDNs that we synthesize, which have five layers in the PDN stack, we train five U-Nets to predict EM hotspots in each specific layer.  

In our testcases, the metal layers M1 and M2 have 30-50\% of their PDN segments as EM-critical, M5 has less than 7\% of EM-critical segments, and M8 and M9 have zero EM-critical segments in the dataset. As stated in Section~\ref{sec:background-em}, the small number of EM-critical segments in M5 makes it extremely challenging to capture hotspots in this layer due to a severe imbalance in data. This imbalance is addressed by using the aggressive threshold across all layers such that the fraction of the number of EM hotspots in M5 increases from 7 to 15\%. In this way 15\% of M5 is {\it EM-prone} while 7\% is EM-critical. While using an aggressive threshold does increase the number of false positives on the original threshold, it ensures EMEDGe does not miss reporting any EM-prone regions. The flagged regions can be checked with a more accurate detailed analysis on a much smaller scale (a few thousand nodes as compared to millions) than the original problem itself. 
%The application of EMEDGe is in detecting regions with EM-prone segments that can later be analyzed in detail using fast solvers. Therefore, EMEDGe which serves as an ML-assisted EM-hotspot identifier and is consistent with the industry requirements where a fast ML inference can quickly identify EM-prone PDN segments and an accurate yet fast small scale simulation can evaluate the predicted hotspots.
%\redfn{I think this has to be motivated earlier in the paper and presented a bit differently. Let's discuss when we meet. \blueHL{I added a paragraph earlier. Let me know if this is sufficient.}}

\subsection{Model Training}
\label{sec:training}
\noindent
We split the data in each set for training, validation, and test sets for each of the models that we train.  The training dataset is
normalized  by subtracting the mean and dividing by the standard deviation and is used to train the network using an ADAM
optimizer~\cite{adam} where the loss function is a pixel-wise mean square error
(MSE).  The convolutional operation in the encoder and the transpose
convolution in the decoder are each followed by ReLU activation to add
non-linearity and L2 regularization to prevent over fitting. The hyperparameters used for training the models are shown in Table~\ref{tbl:train-parameters}.

\begin{table}[]
\centering
\caption{ThermEDGe, IREDGe, and EMEDGe training hyperparameters}
\label{tbl:train-parameters}
\resizebox{0.7\linewidth}{!}{%
\begin{tabular}{|l|r|c|c|r|c|}
\hline
\multicolumn{1}{|c|}{\textbf{\begin{tabular}[c]{@{}c@{}}Training \\ hyperparameters\end{tabular}}} & \multicolumn{1}{c|}{\textbf{\begin{tabular}[c]{@{}c@{}}Static\\ ThermEDGe\end{tabular}}} & \textbf{\begin{tabular}[c]{@{}c@{}}Transient   \\ ThermEDGe\end{tabular}} & \textbf{\begin{tabular}[c]{@{}c@{}}Static   \\ IREDGe\end{tabular}} & \multicolumn{1}{c|}{\textbf{\begin{tabular}[c]{@{}c@{}}Transient \\ IREDGe\end{tabular}}} & \textbf{\begin{tabular}[c]{@{}c@{}}EMEDGe\\  (all models)\end{tabular}} \\ \hline
Loss function & \multicolumn{4}{c|}{Pixelwise MSE} & Cross Entropy Loss \\ \hline
Learning rate & \multicolumn{3}{c|}{1.00E-03} & \multicolumn{2}{c|}{1.00E-04} \\ \hline
Decay rate & \multicolumn{5}{c|}{0.98} \\ \hline
Decay steps & \multicolumn{5}{c|}{1000} \\ \hline
Regularizer & \multicolumn{5}{c|}{L2} \\ \hline
Regularization rate & \multicolumn{3}{c|}{1.00E-05} & \multicolumn{2}{c|}{1.00E-06} \\ \hline
Optimizer & \multicolumn{5}{c|}{ADAM} \\ \hline
Epochs & 500 & \multicolumn{1}{r|}{200} & \multicolumn{1}{c|}{500} & 200 & \multicolumn{1}{c|}{200} \\ \hline
\end{tabular}%
}
\end{table}

\begin{table}
\centering
\caption{ThermEDGe, IREDGe, and EMEDGe ML model parameters}
\label{tbl:layer-parameters}
\resizebox{0.8\linewidth}{!}{%
\begin{tabular}{||l|l||r|r||r|r||r||} 
\hhline{|t:==:t:==:t:==:t:=:t|}
\multicolumn{2}{||c||}{\textbf{Layer hyperparameters}} & \multicolumn{1}{c|}{\begin{tabular}[c]{@{}c@{}}\textbf{Static }\\\textbf{ThermEDGe}\end{tabular}} & \multicolumn{1}{c||}{\begin{tabular}[c]{@{}c@{}}\textbf{Transient}\\\textbf{~ThermEDGe}\end{tabular}} & \multicolumn{1}{c|}{\begin{tabular}[c]{@{}c@{}}\textbf{Static}\\\textbf{~IREDGe}\end{tabular}} & \multicolumn{1}{c||}{\begin{tabular}[c]{@{}c@{}}\textbf{Transient }\\\textbf{IREDGe}\end{tabular}} & \multicolumn{1}{c||}{\begin{tabular}[c]{@{}c@{}}\textbf{EMEDGe }\\\textbf{(all models)}\end{tabular}} \\ 
\hhline{|:==::==::==::=:|}
conv1 & filter size & 5 $ \times $ 5 & 5 $\times$ 5 & 3 $\times$ 3 & 3 $\times$ 3 $\times$ 3 & 3 $\times$ 3 \\ 
\hline
conv\_trans1 & \# filters & 64 & 64 & 64 & 32 & 32 \\ 
\hline
conv2 & filter size & 3 $ \times $ 3 & 3 $ \times $3 & 3 $ \times $3 & 3 $\times $3 $ \times $ 3 & 3 $\times$ 3 \\ 
\hline
conv\_trans1 & \# filters & 32 & 32 & 32 & 32 & 32 \\ 
\hline
conv3 & filter size & 3 $ \times $ 3 & - & 3 $ \times $ 3 & 3 $\times $ 3 $\times $ 3 & 3 $ \times $ 3 \\ 
\hline
conv\_trans3 & \# filters & 16 & - & 16 & 64 & 64 \\ 
\hline
conv4 & filter size & - & - & - & 3 $\times $ 3 $\times $ 3 & 3 $\times $ 3 \\ 
\hline
conv\_trans4 & \# filters & - & - & - & 128 & 128 \\ 
\hline
\multirow{2}{*}{ConvLSTM} & filter size & - & 7 $ \times $ 7 & - & - & - \\ 
\cline{2-7}
 & \# filters & - & 16 & - & - & - \\ 
\hline
\multicolumn{2}{||l||}{Max
  pool layer filters} & 2 $\times$ 2 & 2 $ \times $ 2 & 2 $ \times $ 2 & 2 $ \times $ 2 & 2 $\times$ 2 \\ 
\hhline{|:==::==::==::=:|}
\multicolumn{2}{||l||}{\#Trainable
  parameters} & 132,769 & 235,521 & 133,921 & 458,337 & 261,089 \\
\hhline{|b:==:b:==:b:==:b:=:b|}
\end{tabular}
}
\end{table}

The general architecture for the models used in static ThermEDGe, IREDGe, and EMEDGe is U-Net based and is highlighted in Fig.~\ref{fig:thermedge}, and architecture of transient ThermEDGe is highlighted in Fig.~\ref{fig:lstm-edge}. While the general architecture between all the U-Net-based models is similar, there are some differences. The key differences lie in the convolution layer type (2D vs. 3D), number of layers, and filter sizes because the size of the hotspots (IR drop, or EM, or temperature) are different. These differences between these models are listed in Table~\ref{tbl:layer-parameters}.  The table also lists the number of trainable parameters in each model. It is important to note that the number of parameters in the model is independent of the size of the chip but scales with the size of the hotspot, which is generally similar for a given application domain, technology, and packaging choice. A change in hotspot size or resolution demands a change in the number of layers and receptive field sizes of the models to accurately capture temperature, IR drop, and EM hotspots.
  
The models are trained in Tensorflow 2.1 on an NVIDIA GeForce RTX2080Ti GPU. Training runtimes are as follows: 30m each for static ThermEDGe and IREDGe, 6.5h for transient ThermEDGe, 5 hours for transient IREDGe, and 75 mins for each U-Net model in EMEDGe. 
We reiterate that this is
a one-time cost for a given technology node
and package which is amortized over
repeated use over many design iterations for multiple chips.

%\section{Temperature/IR drop Results using ThermEDGe/IREDGe}
\section{Evaluation of EDGe Networks}
\label{sec:results}
\subsection{Experimental Setup and Metric Definitions}
\noindent
ThermEDGe, IREDGe, and EMEDGe are implemented using Python3.7 within a PyTorch 1.6
framework.  We test the performance of our models on 40 datapoints
reserved in the testset (Section~\ref{sec:training}), labeled T1--T40.  As mentioned in Section~\ref{sec:train}, due to the unavailability of new, public domain benchmarks to evaluate our experiments, 
we use benchmarks that represent commercial industry design workloads. 

\noindent
{\bf Error Metrics} As a measure of goodness of ThermEDGe and IREDGe predictions, we define a discretized regionwise error, $T_{err}~=~\left | T_{true} - T_{pred} \right | $, where $T_{true}$ comes from ground truth image, generated by commercial tools, and $T_{pred}$ from the predicted image, generated by ThermEDGe; $IR_{err}$ is computed similarly. We report the average and maximum values of $T_{err}$ and $IR_{err}$ for each testcase. In addition, the percentage mean and the maximum error is listed as a fraction of a temperature corner, i.e., 105$^\circ$C for thermal analysis and as a fraction of VDD$=0.7$V for IR drop analysis. To evaluate EMEDGe, we use standard binary classification metrics such as accuracy, F1 scores, and area under the ROC curve (AUROC).

\subsection{Static ThermEDGe Accuracy}

\noindent
A comparison between the commercial tool-generated temperature and the
ThermEDGe-generated temperature map for T1--T5 are listed in
Table~\ref{tbl:thermal-results}.  The runtime of static ThermEDGe for
each of the five testcases of size 34$\times$32, is approximately 1.1ms in
our environment. Across the five testcases (five rows of the table), ThermEDGe has an average $T_{err}$ of 0.63$^\circ$C and a maximum $T_{err}$ of 2.93$^\circ$C. These numbers are a small fraction of the maximum ground truth temperature of these testcases (85 -- 150$^\circ$C).
The fast runtimes imply that our method can be used in the
the inner loop of a thermal optimizer, e.g., to evaluate various chip configurations 
under the same packaging solution (typically chosen early in the
design process).  For such applications, this level of error is very
acceptable.

\begin{table}
\centering
\caption{Results of ThermEDGe across 10 testcases.}
\vspace{-1.0em}
\label{tbl:thermal-results}
\resizebox{0.7\linewidth}{!}{%
\begin{tabular}{||l|r|r||r|r|r||} 
\hhline{|t:===:t:===:t|}
\multicolumn{3}{||c||}{Static ThermEDGe} & \multicolumn{3}{c||}{Transient ThermEDGe} \\ 
\hhline{|:===::===:|}
\textbf{Test}~ & \textbf{Avg. $\bf \it T_{err}$ (C)}  & \textbf{Max $\bf \it T_{err}$ (C)}  & \textbf{Test}~ & \textbf{Avg. $\bf \it T_{err}$ (C)}  & \textbf{Max $\bf \it T_{err}$ (C)}  \\ 
\hhline{|:===::===:|}
T1 & 0.64 (0.61\%) & 2.76 (2.63\%) & T6 & 0.51 (0.49\%) & 5.59 (5.32\%) \\ 
\hline
T2 & 0.63 (0.60\%) & 2.67 (2.54\%) & T7 & 0.58 (0.55\%) & 6.17 (5.88\%) \\ 
\hline
T3 & 0.65 (0.62\%) & 2.93 (2.79\%) & T8 & 0.57 (0.54\%) & 5.83 (5.55\%) \\ 
\hline
T4 & 0.48 (0.46\%) & 2.22 (2.11\%) & T9 & 0.52 (0.50\%) & 6.32 (6.02\%) \\ 
\hline
T5 & 0.75 (0.71\%) & 2.86 (2.72\%) & T10 & 0.56  (0.53\%) & 7.14  (6.80\%) \\
\hhline{|b:===:b:===:b|}
\end{tabular}
}
\vspace{-1.0em}
\end{table}

\begin{figure}[h]
\centering
\includegraphics[width=8cm]{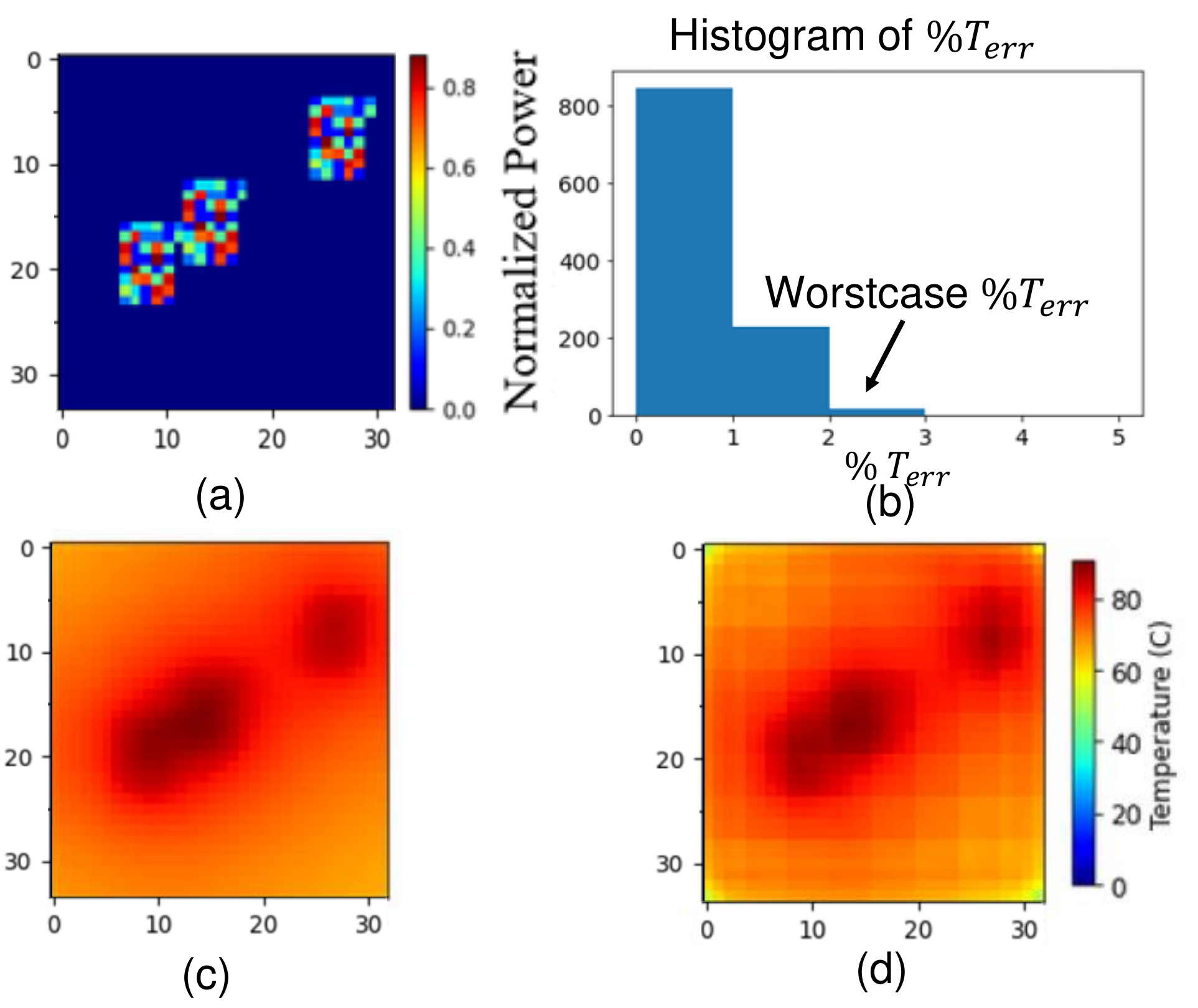}
\caption{Static ThermEDGe temperature estimation on T1: (a) input normalized power distribution, (b) histogram of $T_{err}$, (c) ground truth, and (d) predicted temperature map.}
\label{fig:static-thermal-results}
%\vspace{-1.2em}
\end{figure}

A graphical view of the predicted map for T1 is
depicted in Fig.~\ref{fig:static-thermal-results}. For a given input power
distribution (Fig.~\ref{fig:static-thermal-results}(a)), we compare the true temperature in Fig.~\ref{fig:static-thermal-results}(c) against the ThermEDGe-generated temperature contours plots in
Fig.~\ref{fig:static-thermal-results}(d). The discrepancy is visually seen to
be small. Numerically, the histogram in Fig.~\ref{fig:static-thermal-results}(b) shows the distribution of $T_{err}$ where the worst-case $T_{err}$ is 2.63\% of the temperature corner. 

\subsection{Transient ThermEDGe Accuracy}

\noindent
Transient ThermEDGe predicts the output 200-frame temperature sequence at a 15s interval for the input power sequence. 
We summarize the results in Table~\ref{tbl:thermal-results}.  
Across the five testcases, the prediction has an average $T_{err}$ of 0.52\%
and a maximum $T_{err}$ of 6.80\% as shown. The maximum $T_{err}$ in our testcases occur during transients that do not have long-lasting effects (e.g., on IC reliability).  The magnitude of the maximum $T_{err}$ at sustained peak temperatures is much lower, and is similar to the average $T_{err}$. The inference runtime to generate a 200-frame temperature contour sequence takes 10ms in our setup. In light of these millisecond runtimes these small errors are negligible.

\begin{figure}[ht]
    \centering
    % \animategraphics[width=1.05\linewidth, autoplay, controls]{12}{figs/video-frames/frame}{0}{199}
    \includegraphics[width=0.75\linewidth]{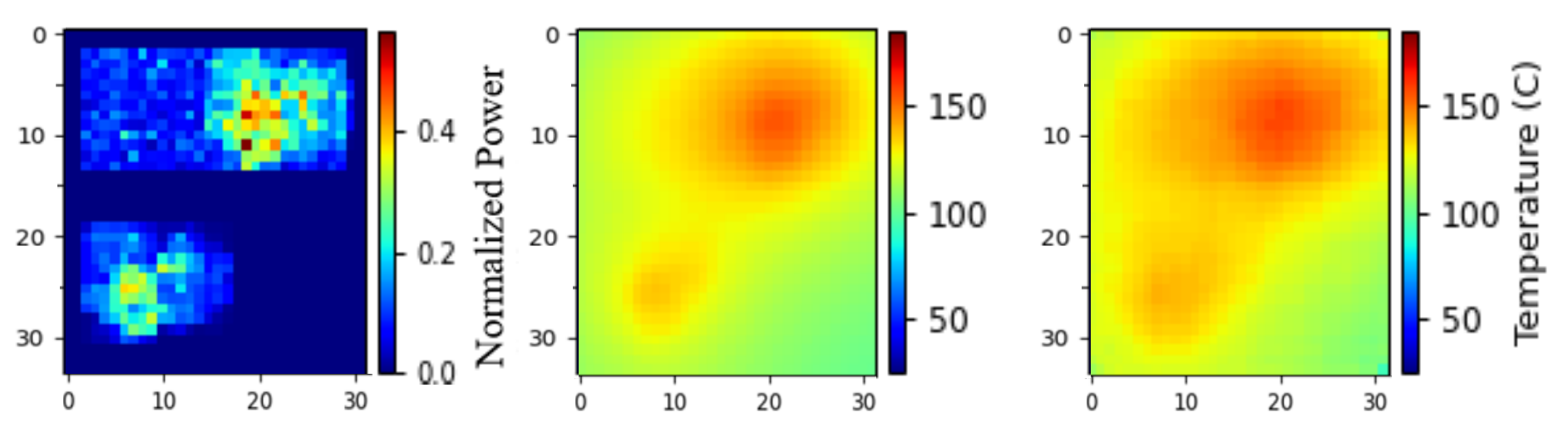}
    \caption{Transient ThermEDGe results on T6 highlighting the input power map (left), commercial tool (center) temperature contours, and transient ThermEDGe predicted IR drop contours at a specific time-step}
    %{\em [For an animated version, visit the GitHub repository~\cite{EDGe-github}]}}
    \label{fig:video}
\end{figure}

Fig.~\ref{fig:video} (left) shows a single frame of a video representing time-varying
power maps for T6, where each frame (time-step) is after a 15s
time interval. The corresponding ground truth and predicted
temperature contours at the current time step are shown in the center and right,
respectively, of the figure. 

% \footnotetext[3]{This media/video can be played on Adobe Acrobat DC Reader Version 9 and above. If this version is unavailable, it can be viewed on this link: https://github.com/asp-dac/asp-dac-1323.git}

\subsection{Static IREDGe Accuracy}
\noindent
We compare IREDGe-generated contours against the contours generated by PDNSim~\cite{pdnsim}. Across the five testcases in Table~\ref{tbl:iredge-qcomm-results}, IREDGe has an average $IR_{err}$ of 0.053mV and a worst-case max $IR_{err}$ of 0.34mV which corresponds to 0.008\% and 0.048\% of VDD respectively. Given that static
IR drop constraints are 1--2.5\% of VDD, a worst-case error of 0.34mV is acceptable in light of rapid runtimes.  We list the  results of the testcases in 
Table~\ref{tbl:iredge-qcomm-results} where the percentage errors in $IR_{err}$ are listed as fraction of VDD$=0.7$V.

\begin{table}[h]
\caption{Results of static IREDGe for 10 different testcases. T16-T20 have a chip size that was not in the training set.}
\centering
\label{tbl:iredge-qcomm-results}
\resizebox{0.7\linewidth}{!}{%
\begin{tabular}{||l|r|r||r|r|r||} 
\hhline{|t:===:t:===:t|}
\multicolumn{3}{||c||}{{\bf Chip size: 34x32
}} & \multicolumn{3}{|c||}{{\bf Chip size: 68x32}} \\  
\hhline{|:===::===:|}
\textbf{}  & \textbf{Avg. $\bf \it IR_{err}$ (mV)} & \textbf{Max $\bf \it IR_{err}$ (mV)} & \textbf{}  & \textbf{Avg. $\bf \it IR_{err}$ (mV)} & \textbf{Max $\bf \it IR_{err}$ (mV)} \\
\hline \hline
T11 &  0.052 (0.007\%)  & 0.26 (0.03\%) & T16 & 0.035 (0.005\%) & 0.16 (0.02\%)\\  \hline
T12 &  0.074 (0.011\%)  & 0.34 (0.05\%) & T17 & 0.054 (0.008\%) & 0.42 (0.06\%)\\ \hline
T13 &  0.036 (0.005\%)  & 0.21 (0.03\%) & T18 & 0.035 (0.005\%) & 0.35 (0.05\%)\\ \hline
T14 &  0.053 (0.008\%)  & 0.24 (0.03\%) & T19 & 0.068 (0.010\%) & 0.22 (0.03\%)\\ \hline
T15 &  0.051 (0.007\%)  & 0.23 (0.03\%) & T20 & 0.061 (0.009\%) & 0.38 (0.05\%)\\
\hhline{|b:===:b:===:b|}
\end{tabular}

}
\end{table}

\begin{figure}[h]
\centering
\includegraphics[width=11cm]{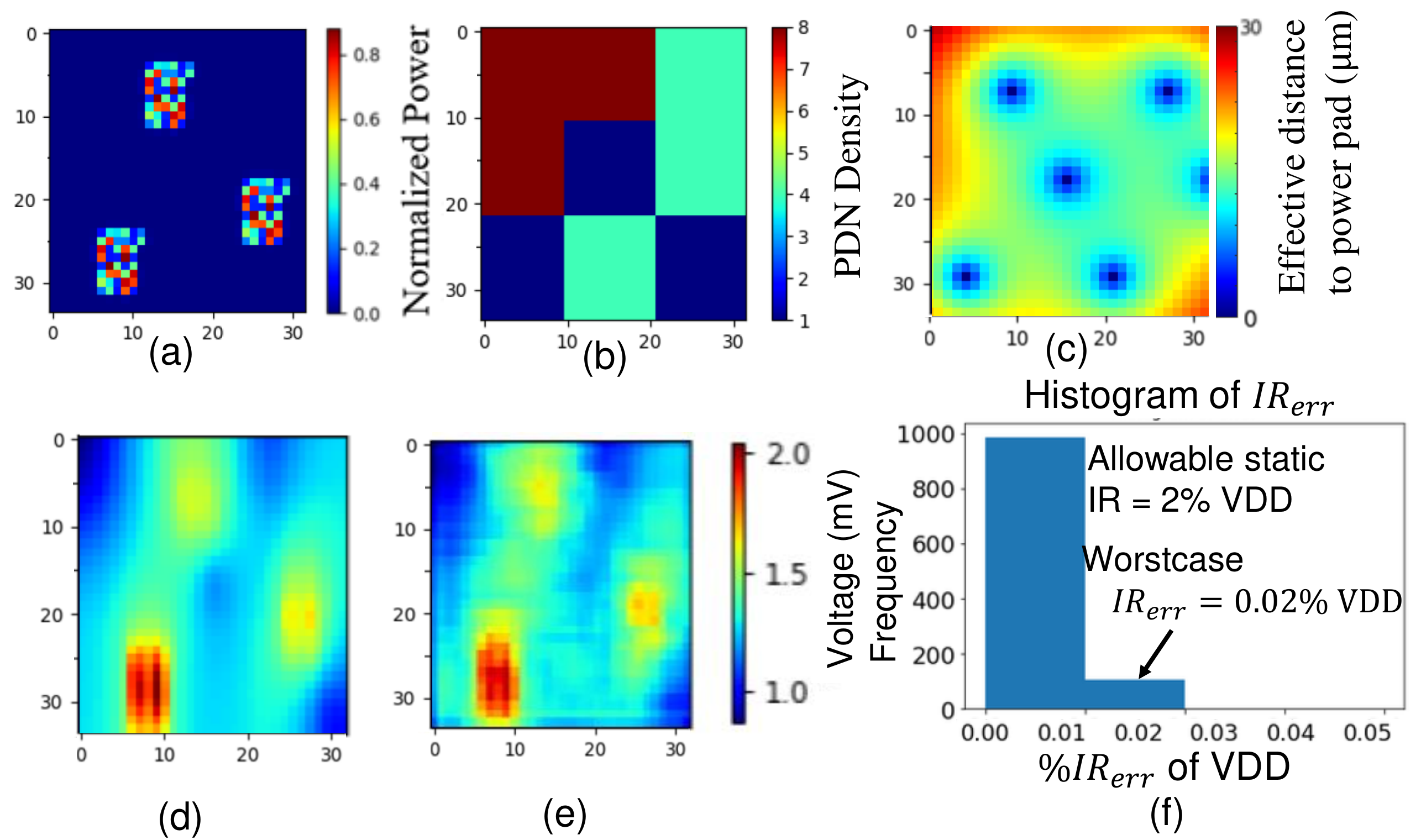}
%\vspace{-2.5em}
\caption{Static IREDGe data for T11: input (a) power map, (b) PDN density map, (c) effective distance map, output (d) ground truth, (e) predicted IR drop map, and (f)  histogram of $IR_{err}$.}
\label{fig:ir-result}
\end{figure}

\begin{figure}[tb]
\centering
\includegraphics[width=8cm]{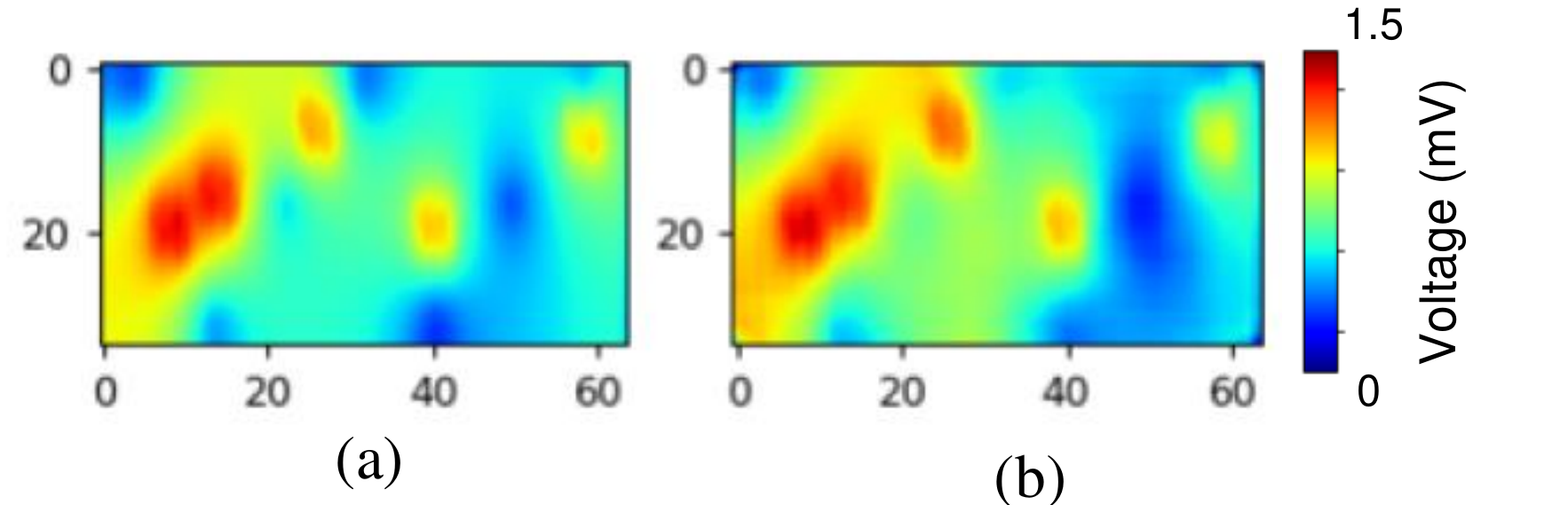}
\vspace{-1.0em}
\caption{Comparison between actual (left) and IREDGe-predicted (right) IR drop contours for images of size 68$\times$32 using a model that 
was trained on images of size 34$\times$32.}
\label{fig:size-independent}
\vspace{-1.0em}
\end{figure}

A detailed view of T11 is shown in Fig.~\ref{fig:ir-result}. 
It compares the IREDGe-generated IR drop contour plots
against contour plot generated by~\cite{pdnsim}.
The input power maps, PDN density maps, and effective distance to power pad maps are shown
in Fig.~\ref{fig:ir-result}(a), (b), and (c) respectively.
Fig.~\ref{fig:ir-result}(d) and (e) shows the comparison between ground truth and predicted value for the corresponding inputs. It is evident that the plots are similar; numerically, the histogram in Fig.~\ref{fig:ir-result}(f)
shows the $IR_{err}$ where the worst $IR_{err}$ is less than 0.02\% of VDD.

\noindent
\underline{\it Size-independence:} Since the EDGe models only comprise the trained weights of the convolutional kernels, the same model can be reused to predict IR drop or temperature contours of a chip of a different size.  We demonstrate this using IREDGe on chips of a different size (T16 -- T20), using an input
power distribution of size $68\times32$.
Fig.~\ref{fig:size-independent} compares the actual IR drop of T16 (left) and the IREDGe-predicted (right) using a model which was trained on $34\times32$ power maps. The results on T16 -- T20 are in Table~\ref{tbl:iredge-qcomm-results}.

\begin{table}
\centering
\caption{Results of  transient IREDGe on testcases T21-T25.}
\label{tbl:dynamic-ir}
\resizebox{0.4\linewidth}{!}{%
\begin{tabular}{||l|l|l||} 
\hhline{|t:===:t|}
{\bf Test} & {\bf Avg.~$IR_{err}$ (mV)} & {\bf Max~$IR_{err}$ (mV)} \\ 
\hhline{|:===:|}
T21 & 0.47 (0.07\%) & 2.46 (0.35\%) \\ 
\hline
T22 & 0.43 (0.06\%) & 2.24 (0.32\%) \\ 
\hline
T23 & 0.44 (0.06\%) & 2.15 (0.31\%) \\ 
\hline
T24 & 0.3 (0.04\%) & 2.36 (0.34\%) \\ 
\hline
T25 & 0.58 (0.08\%) & 2.03 (0.29\%) \\ 
\hhline{|b|===:b|}
\end{tabular}
}
\end{table}

\begin{figure}[h]
\centering
\includegraphics[width=11cm]{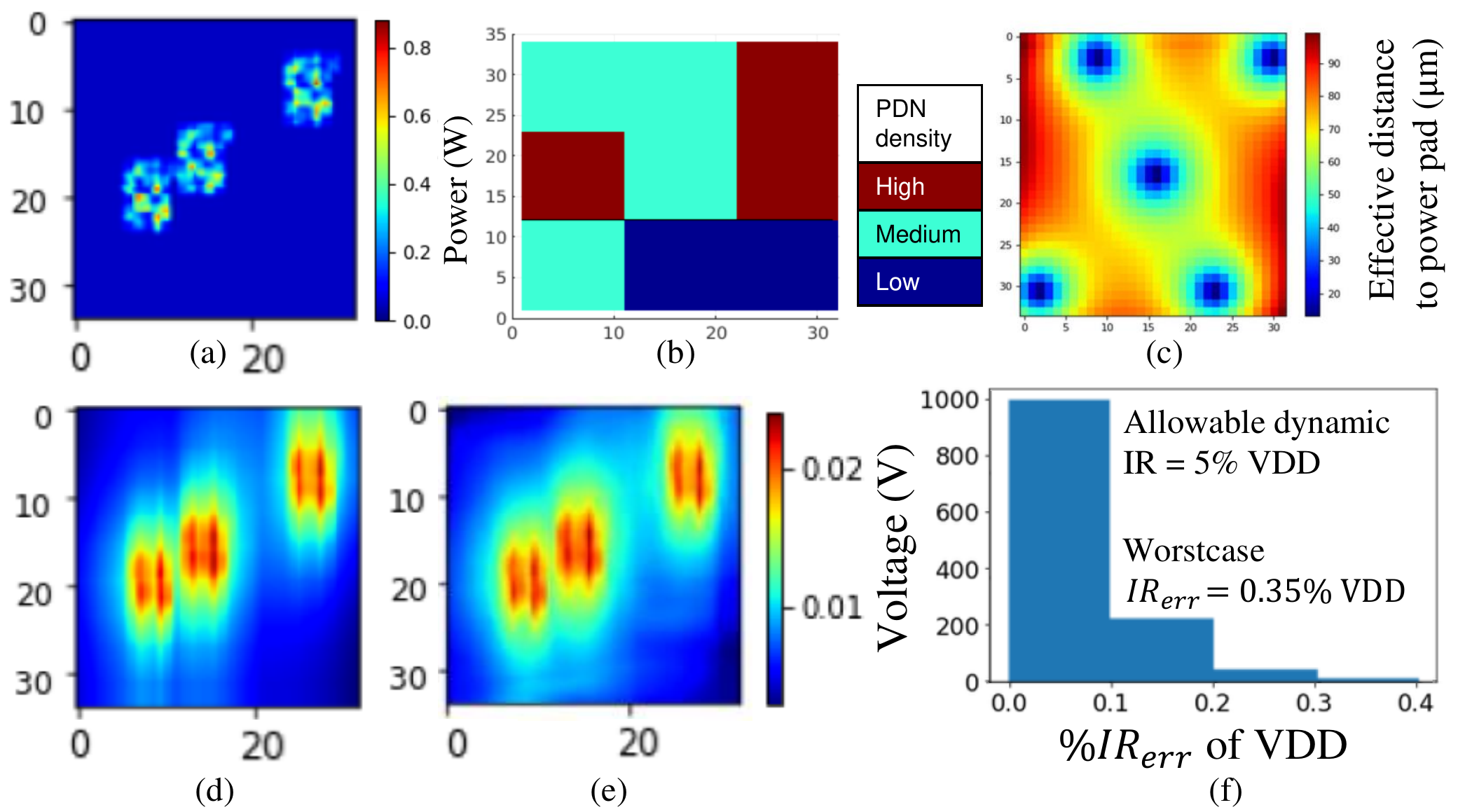}
%\vspace{-2.5em}
\caption{Transient IREDGe data for T21: input (a)~power map at time $t=1$ns, (b)~PDN density map, (c)~effective distance map, output (d)~ground truth, (e)~predicted worst-case dynamic IR drop map for time period $0$ to $2$ns, and (f)  histogram of $IR_{err}$.}
\label{fig:dynamic-ir-result}
\end{figure}

\subsection{Transient IREDGe Accuracy}

\noindent
We compare transient IREDGe against the IR drop contours generated by the ground-truth solver. Table~\ref{tbl:dynamic-ir} shows the results of five testcases. The average $IR_{err}$ and  max $IR_{err}$ is  0.58mV and 3.03mV which corresponds to 0.08\% and 0.43\% of VDD respectively.  Compared to industry-standard transient IR drop limits which can be as high as 5--10\% of VDD, these errors are minimal. 

A detailed result for T21 is shown in Fig.~\ref{fig:dynamic-ir-result}. The inputs including the power map (at $t=1$ns), the PDN density map, and the effective-distance-to-power-pads map, are shown in Fig.~\ref{fig:dynamic-ir-result}(a), (b), and (c) respectively. The ground-truth worst IR drop solution for the entire 2ns period is shown in Fig.~\ref{fig:dynamic-ir-result}(d) and is compared against the IREDGe predicted solution (maximum of the 20 individual inferences as shown by the red line in Fig.~\ref{fig:trans-ir-time-steps}) in Fig.~\ref{fig:dynamic-ir-result}(e). Visually, the predicted IR contours look similar to the ground-truth contours capturing the three hotspots. The histogram in Fig.~\ref{fig:dynamic-ir-result}(f) represents the $IR_{err}$ as a fraction of VDD for all pixels in the IR drop map. This testcases has a worst-case $IR_{err}$ of 0.35\% of VDD with an IR drop limit of 5\% of VDD.

% \begin{table*}[t]
% \centering
% \caption{EMEDGe results for binary hotspot classification on a per-PDN segment basis for testcases T26-T30.}
% \resizebox{0.8\linewidth}{!}{%
% \begin{tabular}{||l|r|l|l|l|l|r||l|l|l|l|l|l||l|l|l|l|l|l||} 
% \hhline{|t:=======:t:======:t:======:t|}
%  & \multicolumn{6}{c||}{M1} & \multicolumn{6}{c||}{M2} & \multicolumn{6}{c||}{M5} \\ 
% \hhline{|:=======::======::======:|}
% Test & \multicolumn{1}{l|}{\begin{tabular}[c]{@{}l@{}}ROC \\AUC\end{tabular}} & TP & TN & FP & FN & \multicolumn{1}{l||}{F1} & \begin{tabular}[c]{@{}l@{}}ROC \\AUC\end{tabular} & TP & TN & FP & FN & F1~ & \begin{tabular}[c]{@{}l@{}}ROC \\AUC\end{tabular} & TP & TN & FP & FN & F1~ \\ 
% \hhline{|:=======::======::======:|}
% T26 & 0.99 & 65,897 &205,395 & 9,963 & 2,645 & 0.83 &  &  &  &  &  &  &  &  &  &  &  &  \\ 
% \hline
% T27 & 0.99 & 138,395 & 133,100 & 6,195 & 6,210 & 0.91 &  &  &  &  &  &  &  &  &  &  &  &  \\ 
% \hline
% T28 &  &  &  &  &  &  &  &  &  &  &  &  &  &  &  &  &  &  \\ 
% \hline
% T29 & 0.99 & 245,779 & 32,000 & 24,50 & 3,671 & 0.83 &  &  &  &  &  &  &  &  &  &  &  &  \\ 
% \hline
% T30 &  &  &  &  &  &  &  &  &  &  &  &  &  &  &  &  &  &  \\
% \hhline{|b:=======:b:======:b:======:b|}
% \end{tabular}
% }
% \end{table*}

\subsection{IREDGe Compared with PowerNet}
\noindent
We implemented a version of PowerNet~\cite{powernet} for both static and transient IR drop analysis. For both analysis, the layout is divided into tiles, and the CNN features are the
2-D power distributions (toggle rate-scaled switching and internal power, total
power, and leakage power) within each tile and in a fixed window of surrounding tiles.
The trained CNN predicts the IR drop on a tile-by-tile basis by
sliding a window across all tiles on the chip. The work uses a tile size of
5$\mu$m$\times$5$\mu$m and takes into consideration a 31$\times$31 tiled
neighborhood (window) power information as features. 
For a fair comparison, we train IREDGe under a fixed PDN density and fixed power pad locations used to train PowerNet.  
For transient analysis, PowerNet uses a maximum CNN structure which takes the power map at the current time-step and uses it to predict a single value of IR drop for a specified tile at the current time-step. To generate the wortcase IR drop map, the maximum CNN structure finds the maximum IR drop of all time-steps for each tile on the chip. For static analysus, PowerNet leverages a CNN and predicts a single IR drop value per tile.

\noindent
Qualitatively, IREDGe is superior to PowerNet in three aspects: \\
(i) {\em Tile and Window Size Selection:} It is stated in~\cite{powernet} that 
when the size of the tile is increased from
1$\mu$m$\times$1$\mu$m to 5$\mu$m$\times$5$\mu$m and the size of the resulting window is increased
to represent 31$\times$31 window of 25$\mu$m$^2$ tiles instead of 1$\mu$m$^2$ tiles, the 
accuracy of the PowerNet model improves. 
In general, this is the
expected behavior with an IR analysis problem where the accuracy increases as
more global information is available, until a certain radius after which the
principle of locality  holds~\cite{Chiprout04}. IREDGe bypasses
this tile-size selection problem entirely by providing the entire power map
as input to IREDGe and allowing the network to learn the window size needed for accurate IR estimation.  \\
(ii) {\em Runtimes and accuracy:} 
We compare IREDGe against our implementation of PowerNet on T26--35. The first five of these ten testcases are for static IR drop, while the latter five are for transient IR drop. 
These testcases have identical power distributions in T11--15 and T21--25 except that all
the ten testscases have identical uniform PDNs and identical power pad
distributions, as required by PowerNet; IREDGe does not require this. 

\begin{figure}[h]
\centering
\includegraphics[width=9cm]{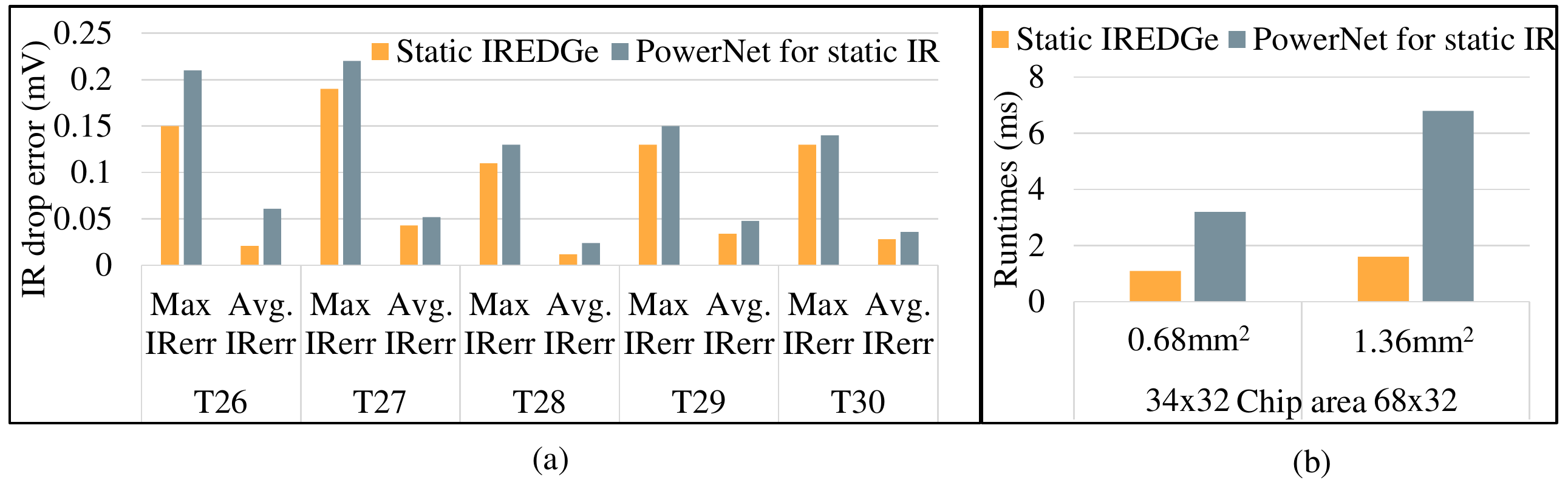}
\vspace{-1.0em}
\caption{Static IREDGe versus PowerNet: (a) IR drop error and (b) runtimes. IREDGe is 2.9$\times$ faster at iso-error across T26--T30 (0.68mm$^2$ area).}
\vspace{-1.0em}
\label{fig:compare-powernet-iso-accuracy}
\end{figure}

Static IREDGe requires a {\em single} inference, irrespective of the size of the chip, while the static version of PowerNet performs an inference for every tile in the chip as it predicts IR drop by sliding a window on a tile-by-tile basis. In our experimental setup, it takes 75 minutes to train PowerNet, as against 30 minutes for IREDGe. Fig.~\ref{fig:compare-powernet-iso-accuracy} shows the comparison of inference times between PowerNet and IREDGe at similar error levels. At the 25$\mu$m$^2$ tile size, both IREDGe and PowerNet have similar accuracies for T26--30 as shown in Fig.~\ref{fig:compare-powernet-iso-accuracy}(a). At this error level, for 0.68 mm$^2$ designs, IREDGe is  2.9$\times$ faster than PowerNet (Fig.~\ref{fig:compare-powernet-iso-accuracy}(b)). 

\begin{figure}[h]
\centering
\includegraphics[width=9cm]{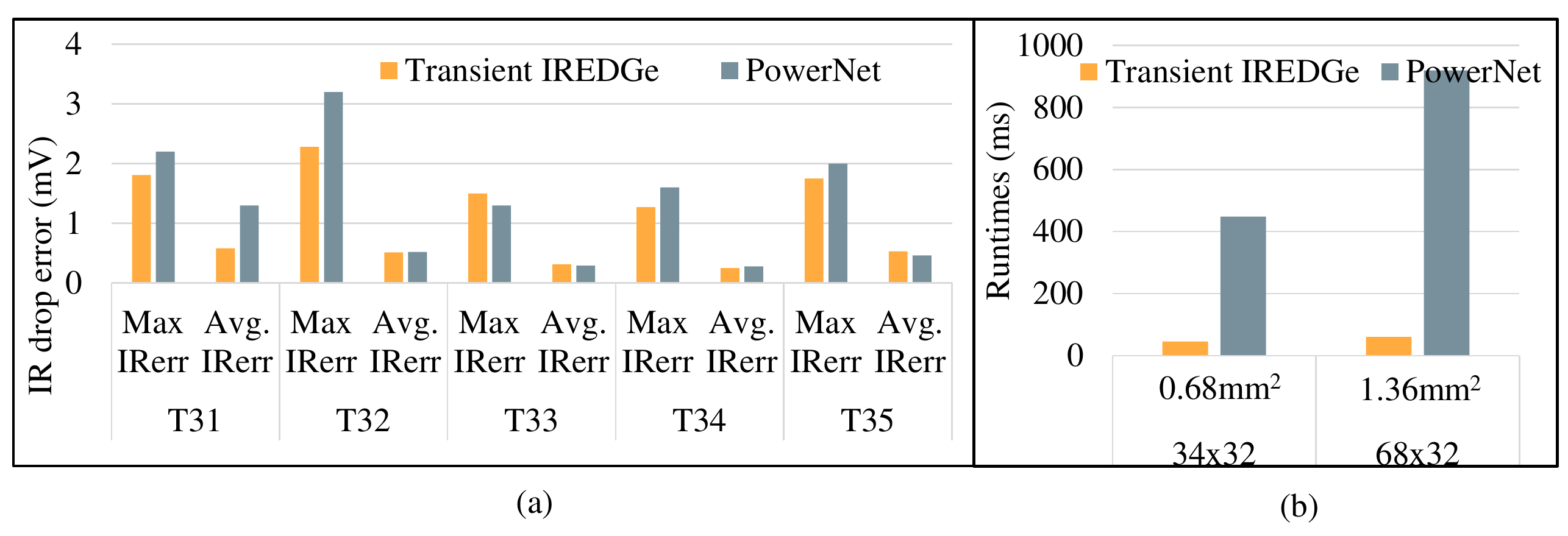}
\vspace{-1.0em}
\caption{Transient IREDGe versus PowerNet: (a) IR drop error and (b) runtimes. IREDGe is 9.7$\times$ (0.68mm$^2$) faster at iso-error across T31--T35.}
\vspace{-1.0em}
\label{fig:powernet-comp-trans}
\end{figure}

The inference time for transient IREDGe is higher than static IREDGe as it requires twenty inferences with 3D convolutional layers and takes ten power maps at different time steps as input compared to static IREDGe that uses 2D convolutional layers with a single power map as input. However, transient IREDGe is still faster than PowerNet as it requires a single inference in space and 20 inferences (Refer Section~\ref{sec:train}) in time while PowerNet requires $34 \times 32$ inferences across space and $100$ across time for the $100$ time steps in a 0.2ns clock period. IREDGe can operate at a coarser granularity due to the 3D convolutional layers that capture time-dependent effects, unlike the CNN in PowerNet that performs an inference for each time step. The errors of transient IREDGe are compared against PowerNet and shows in Fig.~\ref{fig:powernet-comp-trans}(a), and the runtime comparisons for transient IR drop are listed in Fig.~\ref{fig:powernet-comp-trans}(b). Similar to static IREDGe, for iso-accuracy (Fig.~\ref{fig:powernet-comp-trans}(a)) has transient IREDGe is 9.7$\times$ faster than PowerNet for a chip area of 0.68mm$^2$ and 15.3$\times$ faster for a chip of area 1.36mm$^2$.

%Across T21--25 IREDGe has an average $IR_{err}$ of 0.028mV and a maximum $IR_{err}$ of 0.14mV as against 0.044mV and 0.17mV respectively for PowerNet as shown in Fig.~\ref{fig:compare-powernet-iso-accuracy} (left). 

%For inference, PowerNet takes 3.2ms while IREDGe takes 1.1ms for a chip area of 68mm$^2$ which corresponds to a speedup of 2.9$\times$ and a speedup of 5.6$\times$ for a chip area of 136mm$^2$ (Fig.~\ref{fig:compare-powernet-iso-accuracy}(right)) with IREDGe having a slightly lower error (Fig.~\ref{fig:compare-powernet-iso-accuracy}(left)). 

\begin{figure}[h]
\centering
\includegraphics[width=9cm]{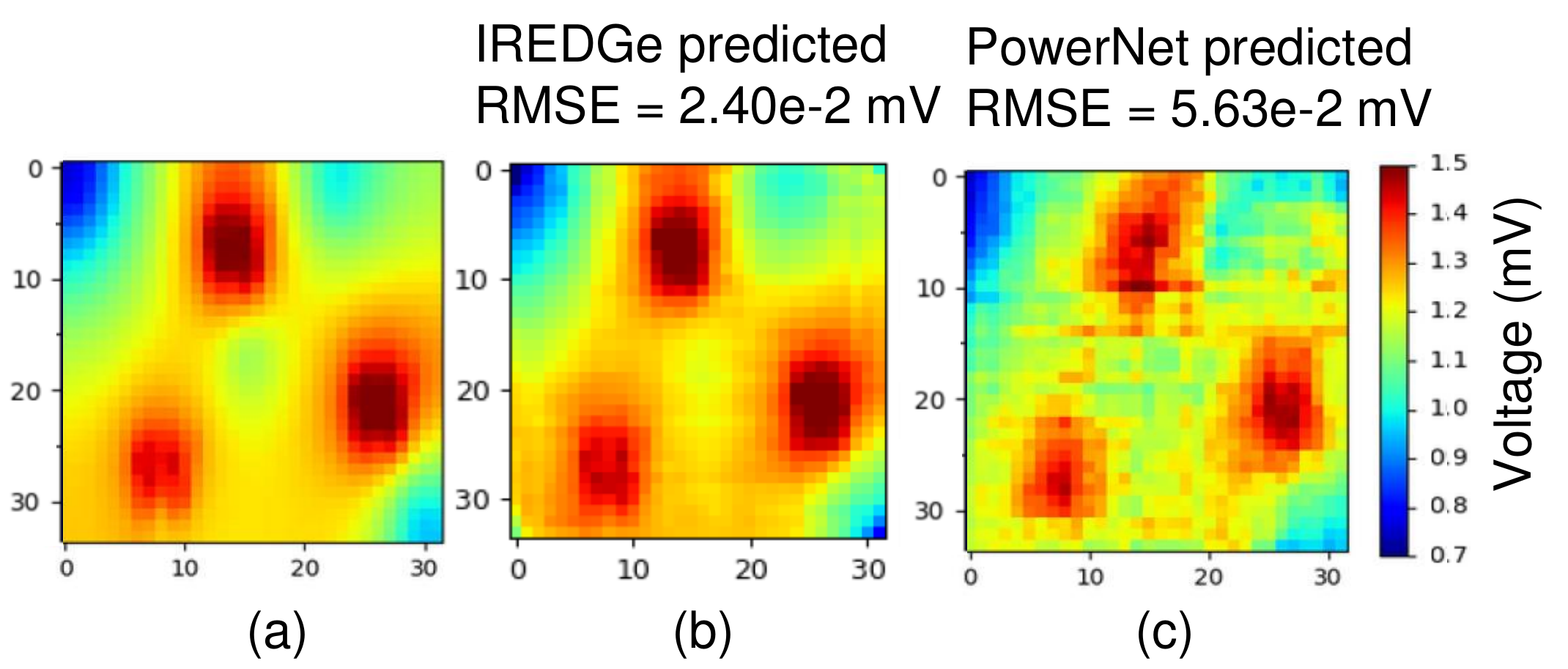}
\caption{IR drop comparisons on T26: (a) ground truth, (b) static IREDGe, and (c) our implementation of static PowerNet.}
\label{fig:powernet-comp}
\end{figure}

\noindent
(iii) {\em Pixelated IR drop maps:} Since PowerNet uses a CNN to predict IR drop on a
tile-by-tile basis, where each tile is 5$\mu$m $\times$ 5$\mu$m,
the resulting IR drop image is pixelated, and the predicted region value does not correlate well with the neighboring regions.  This is highlighted in
Fig.~\ref{fig:powernet-comp} which compares IR drop contours
from a golden solver (Fig.~\ref{fig:powernet-comp}(a)), IREDGe
(Fig.~\ref{fig:powernet-comp}(b)), and our implementation of
PowerNet (Fig.~\ref{fig:powernet-comp}(c)) for T26.

\subsection{Impact of C4 bump locations and PDN density on IR drop}
\label{sec:C4bump_PDNdensity}
\begin{figure}[t]
\centering
\includegraphics[width=14cm]{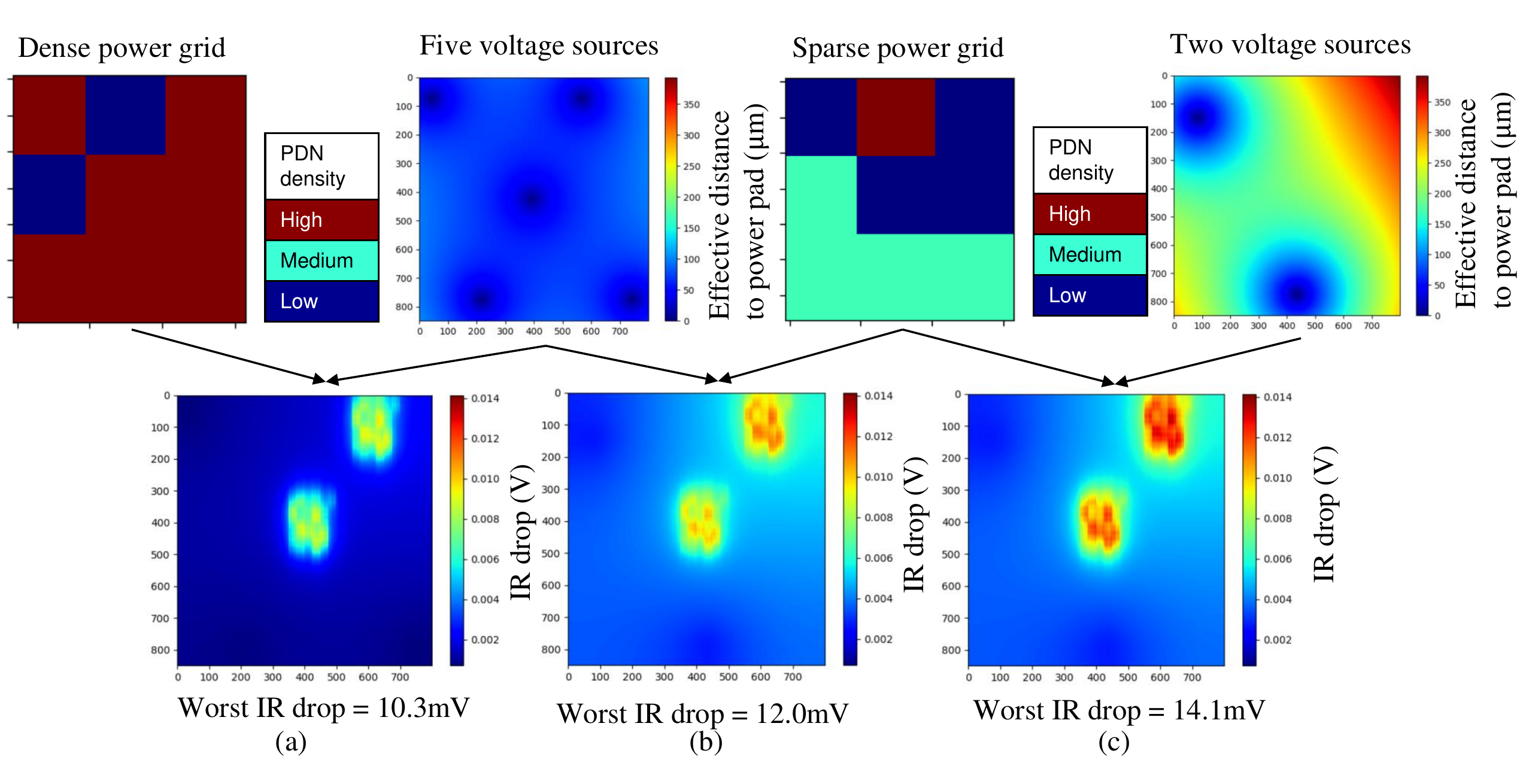}
\caption{Comparison of worst-case transient IR drop distributions for (a) a dense PDN with five voltage sources, (b) a sparse PDN with five voltage sources, and (c) a sparse PDN with only two voltage sources for a fixed input power map.}
\label{fig:impact-of-vsrc-pdn}
\end{figure}

In addition to the on-chip power distributions, the locations of the C4 bumps and PDN densities dictate both static and transient IR drop distributions. The voltage source locations and PDN density impact the equivalent resistance between the on-chip instance and the power supply affecting the IR drop pattern.
However, the works in~\cite{MAVIREC_date21, powernet} are based on input power only and do not account for these additional factors that are critical for accurate IR drop prediction. We measure the impact these factors have on worst-case transient IR drop in Fig.~\ref{fig:impact-of-vsrc-pdn} and highlight the importance of considering these factors as features in IREDGe.  The figure shows two configurations of PDN densities and two configurations of voltage source locations for a fixed input power distribution. The worst-case transient IR drop in Fig.~\ref{fig:impact-of-vsrc-pdn}(a) is 14.2\% lower than worst IR drop in Fig.~\ref{fig:impact-of-vsrc-pdn}(b) due to the higher PDN density in the former. The impact of the C4 bump distribution on IR drop is highlighted in Fig.~\ref{fig:impact-of-vsrc-pdn}(b) and (c). For the same power grid and input power distributions, the worst-case IR drop changes by 14.9\% due to a change in the number and locations of the C4 bumps. Therefore, capturing these degrees of freedom is critical to accurate IR drop prediction.

Table~\ref{tbl:impact-of-vsrc-pdn} compares max $IR_{err}$ and average $IR_{err}$ of transient IREDGe, against a model such as~\cite{MAVIREC_date21} which does not consider the location of C4 bumps and PDN density as features. The table shows that for testcases T21--T25 these features are required for accurate IR drop prediction. 

\begin{table}
\centering
\caption{Impact of C4 bump and PDN density on testcases T21-T25.}
\label{tbl:impact-of-vsrc-pdn}
\resizebox{0.75\linewidth}{!}{%
\begin{tabular}{||r||r|r||r|r||} 
\hhline{|t:=:t:==:t:==:t|}
\multicolumn{1}{||c||}{\multirow{2}{*}{Test}} & \multicolumn{2}{c||}{\begin{tabular}[c]{@{}c@{}}Transient IREDGe (With~C4 bump location\\~and PDN density as features )\end{tabular}} & \multicolumn{2}{c||}{\begin{tabular}[c]{@{}c@{}}Without C4 bump locations and \\PDN density as features~\end{tabular}} \\ 
\cline{2-5}
\multicolumn{1}{||c||}{} & \multicolumn{1}{c|}{Avg. $IR_{err}$ (mV)} & \multicolumn{1}{c||}{Max $IR_{err}$ (mV)} & \multicolumn{1}{c|}{Avg. $IR_{err}$ (mV)} & \multicolumn{1}{c||}{Max $IR_{err}$(mV)} \\ 
\hhline{|:=::==::==:|}
\multicolumn{1}{||l||}{T21} & 0.47 (0.07\%) & 2.46 (0.35\%) & 0.92 (0.13\%) & 4.78 (0.68\%) \\ 
\hline
T22 & 0.43 (0.06\%) & 2.24 (0.32\%) & 1.37 (0.19\%) & 6.22 (0.88\%) \\ 
\hline
T23 & 0.44 (0.06\%) & 2.15 (0.31\%) & 0.63 (0.09\%) & 3.15 (0.45\%) \\ 
\hline
T24 & 0.3 (0.04\%) & 2.36 (0.34\%) & 0.74 (0.11\%) & 3.89 (0.55\%) \\ 
\hline
T25 & 0.58 (0.08\%) & 2.03 (0.29\%) & 0.67 (0.10\%) & 3.36 (0.48\%) \\
\hhline{|b:=:b:==:b:==:b|}
\end{tabular}
}
\end{table}

\subsection{EMEDGe evaluation}

\noindent
As mentioned in Section~\ref{sec:prob-def} we convert the EM problem into an image-segmentation problem by assigning a pixel as EM-critical if any PDN segment in that region is EM-critical. Since EMEDGe operates on a region level where all the PDN segments in the region are represented by a single pixel, we lose fine-grained information at the PDN-segment level. To fairly evaluate EMEDGe we account for this inaccuracy by  performing comparison against the ground-truth on a per-PDN-segment level. The EMEDGe-predicted EM-prone regions are converted into PDN-segment level detail by assigning all segments in the region to the same class (EM-prone or not) as the pixel.

\begin{figure}[t]
\centering
\includegraphics[width=9cm]{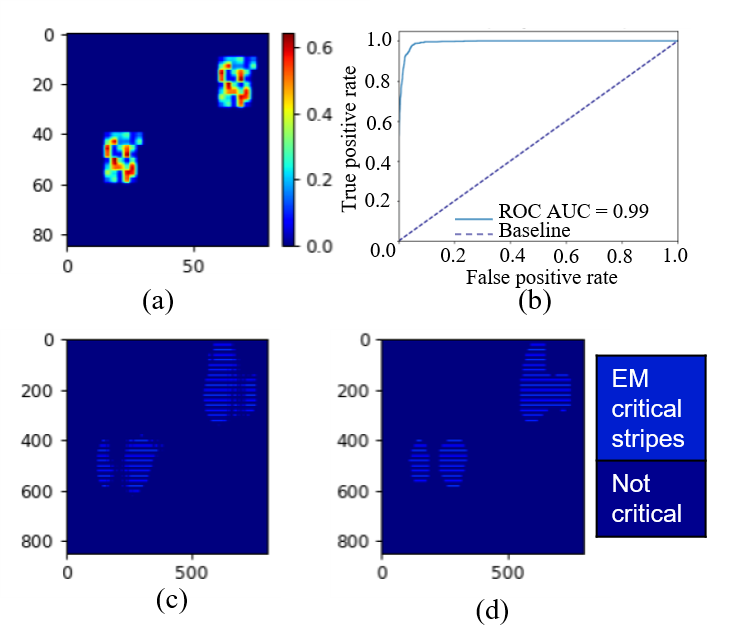}
%\vspace{-2.5em}
\caption{EMEDGe evaluation on T36 for layer M1: input (a)~input chip power map, (b)~ROC curve for on a per-PDN-segment basis, (c)~output ground truth EM-critical PDN segments, and (d)~predicted EM-prone PDN segments.}
\label{fig:em-result}
\end{figure}

\begin{table*}
\centering
\caption{EMEDGe results for binary hotspot classification with original thresholds  on a per-PDN segment basis for testcases T26-T30.}
\label{tbl:em-result-original-threshold}
\resizebox{\linewidth}{!}{%
\begin{tabular}{||c||r|r|r|r|r|r||r|r|r|r|r|r||r|r|r|r|r|r||} 
\hhline{|t:=:t:======:t:======:t:======:t|}
\multirow{2}{*}{\textbf{Test}} & \multicolumn{6}{c||}{\textbf{M1}} & \multicolumn{6}{c||}{\textbf{M2}} & \multicolumn{6}{c||}{\textbf{M5}} \\ 
\cline{2-19}
 & \multicolumn{1}{c|}{\textbf{Accuracy}} & \multicolumn{1}{c|}{\textbf{TP}} & \multicolumn{1}{c|}{\textbf{TN}} & \multicolumn{1}{c|}{\textbf{FP}} & \multicolumn{1}{c|}{\textbf{FN}} & \multicolumn{1}{c||}{\textbf{F1}} & \multicolumn{1}{c|}{\textbf{Accuracy}} & \multicolumn{1}{c|}{\textbf{TP}} & \multicolumn{1}{c|}{\textbf{TN}} & \multicolumn{1}{c|}{\textbf{FP}} & \multicolumn{1}{c|}{\textbf{FN}} & \multicolumn{1}{c||}{\textbf{F1}} & \multicolumn{1}{c|}{\textbf{Accuracy}} & \multicolumn{1}{c|}{\textbf{TP}} & \multicolumn{1}{c|}{\textbf{TN}} & \multicolumn{1}{c|}{\textbf{FP}} & \multicolumn{1}{c|}{\textbf{FN}} & \multicolumn{1}{c||}{\textbf{F1}} \\ 
\hhline{|:=::======::======::======:|}
\textbf{T36} & 0.91 & 116383 & 139666 & 26559 & 0 & 0.90 & 0.92 & 2031 & 5855 & 688 & 0 & 0.86 & 0.98 & 1532 & 31246 & 793 & 0 & 0.79 \\ 
\hline
\textbf{T37} & 0.89 & 138811 & 112144 & 31346 & 0 & 0.90 & 0.86 & 13000 & 15421 & 4534 & 0 & 0.85 & 0.93 & 1433 & 6397 & 579 & 0 & 0.83 \\ 
\hline
\textbf{T38} & 0.88 & 139058 & 108457 & 34840 & 0 & 0.89 & 0.89 & 5717 & 9313 & 1935 & 0 & 0.86 & 0.93 & 4471 & 26736 & 2433 & 0 & 0.79 \\ 
\hline
\textbf{T39} & 0.90 & 138261 & 115856 & 28248 & 0 & 0.91 & 0.93 & 5374 & 10472 & 1156 & 0 & 0.90 & 0.97 & 438 & 7759 & 214 & 0 & 0.80 \\ 
\hline
\textbf{T40} & 0.89 & 112828 & 139866 & 29952 & 0 & 0.88 & 0.92 & 3345 & 12494 & 1389 & 0 & 0.83 & 0.99 & 272 & 8078 & 123 & 0 & 0.82 \\ 
\hhline{|:=::======::======::======:|}
\textbf{\%FP} & \multicolumn{6}{c||}{\textbf{0.106}} & \multicolumn{6}{c||}{\textbf{0.104}} & \multicolumn{6}{c||}{\textbf{0.044}} \\
\hhline{|b:=:b:======:b:======:b:======:b|}
\end{tabular}
}
\end{table*}

\begin{table}
\centering
\caption{Runtimes of EDGe networks and golden analysis tools for a chip of area 0.68mm$^2$.}
\label{tab:runtime}
\resizebox{0.5\linewidth}{!}{%
\begin{tabular}{|l|r|r|r|} 
\hline
\textbf{Analysis type} & \multicolumn{1}{l|}{\textbf{\# Nodes}} & \multicolumn{1}{l|}{\begin{tabular}[c]{@{}l@{}}\textbf{Icepak/}\\\textbf{ PDNSim}\\\textbf{ runtimes}\end{tabular}} & \multicolumn{1}{l|}{\begin{tabular}[c]{@{}l@{}}\textbf{EDGe}\\\textbf{ network }\\\textbf{ runtimes}\end{tabular}} \\ 
\hline
Static thermal & 2.0 million & 30 mins & 1.1ms \\ 
\hline
Transient thermal & 2.0 million & 210 mins & 10ms \\ 
\hline
Static IR drop & 3–8 million & 5–20 mins & 1.1ms \\ 
\hline
Transient IR drop & 3–8 million & 30–60 mins & 46ms \\ 
\hline
EM hotspot & 3–8 million & 5–20 mins & 5.5ms \\
\hline
\end{tabular}
}
\end{table}

Fig.~\ref{fig:em-result} compares the EM-critical segments reported by the ground-truth solver and the EM-prone segments reported by EMEDGe for M1 and an input power map shown in Fig.~\ref{fig:em-result}(a).  Visually, the ground truth EM hotspots from Fig.~\ref{fig:em-result}(c) are near-identical to the EMEDGe-predicted hotspots in Fig.~\ref{fig:em-result}(d). Since the EM hotspot analysis is formulated as a classical binary classification problem, we plot the receiver operating characteristic (ROC) curve, which demonstrates the performance of the model for different thresholds. The area under the ROC (AUROC) curve denotes the accuracy of the model. In this case, for the testcase T26 and the M1 layer model, we have an AUROC of 0.99 as shown in Fig.~\ref{fig:em-result}(b), which shows that the model has a good classification accuracy. 

Even though we perform training with an aggressive threshold compared to the PDK-specified threshold we evaluate the results against PDK-specified original thresholds. 

Table~\ref{tbl:em-result-original-threshold} lists the F1 score, the number of true positives, false positives, true negatives, and false negatives on a per-segment basis for the five testcases across layer M1, M2, and M5 for the scaled aggressive threshold. Even with an aggresive threshold, only the lower metal layers M1, M2, and M5 have EM hotspots, and the other layers do not have any EM-prone segments. Therefore, we do not perform any evaluation on these layers. Our findings match this industry observation where the ratio of EM-critical segments to the total number of segments in each layer reduces from the lower metal layers to the higher metal layers. Across the five testcases listed in Table~\ref{tbl:em-result-original-threshold}, we obtain an average accuracy (average of all rows) of 0.89, 0.9, and 0.96 and average F1 score of 0.90, 0.87, and 0.85 for the models corresponding to layers M1, M2, and M5, respectively, with these aggressive thresholds.

The table shows no false negatives, i.e., EMEDGe has not missed flagging out any EM-critical segment as EM-prone. While there are false positives reported, they are still small in number ($\approx $10\% of the total segments in the PDN). The flagged EM-prone segments (TP and FP) can be further evaluated in detail using a more accurate, fine grained simulation.  The small number of flagged segments implies that such a simulation involves a small subcircuit, and can be fast.

\subsection {Inference Runtime Analysis} 

\noindent
Table~\ref{tab:runtime} compares the inference runtimes of our ML-based EDGe network approach against the ``golden'' solvers. The runtimes are reported on an NVIDIA GeForce RTX 2080Ti GPU.  With the millisecond inference times and the transferable nature of our trained models, the one-time cost of training the EDGe networks is quickly amortized over multiple uses within a design cycle and over multiple designs.

\section{Conclusion}
\noindent
This paper addresses the compute-intensive tasks of thermal and PDN analysis by proposing the use EDGe networks as apt ML-based solutions. 
% Our EDGe-based solution not only improves runtimes over commercial solvers, but overcomes the window size-selection challenge (amount of neighborhood information required for accurate thermal and IR analysis), that is faced by other ML-based techniques, by allowing ML to learn the window size. 
We successfully evaluate EDGe networks for these applications by developing two ML-based tools (i)~ThermEDGe, (ii)~IREDGe, and (iii)~EMEDGe for rapid on-chip thermal, PDN and EM hotspot classification, respectively. The EDGe-based techniques provide millisecond-inference runtimes, and are demonstrated to be accurate when compared to ground-truth solvers, which take several hours.
\bibliographystyle{acm-reference}
\bibliography{references2}

\end{document}